\documentclass[pra,twocolumn,longbibliography,10pt,aps,floatfix,superscriptaddress]{revtex4-1}

\usepackage{array}
\usepackage{graphicx}
\usepackage{verbatim}
\usepackage{color} 
\usepackage{amsfonts}
\usepackage{amsmath}
\usepackage{fancyhdr}

\newcommand{\ket}[1]{\left\vert{#1}\right\rangle}

\renewcommand{\vec}[1]{\mathbf{#1}}

\begin{document}

\title{A logical qubit in a linear array of semiconductor quantum dots}
\author{Cody Jones}
\email{ncodyjones@gmail.com}
\affiliation{HRL Laboratories, LLC, 3011 Malibu Canyon Road, Malibu, CA 90265, USA}
\author{Michael A. Fogarty}
\affiliation{Centre for Quantum Computation and Communication Technology, School of Electrical Engineering and Telecommunications, The University of New South Wales, Sydney, New South Wales 2052, Australia}
\author{Andrea Morello}
\affiliation{Centre for Quantum Computation and Communication Technology, School of Electrical Engineering and Telecommunications, The University of New South Wales, Sydney, New South Wales 2052, Australia}
\author{Mark F. Gyure}
\affiliation{HRL Laboratories, LLC, 3011 Malibu Canyon Road, Malibu, CA 90265, USA}
\author{Andrew S. Dzurak}
\affiliation{Centre for Quantum Computation and Communication Technology, School of Electrical Engineering and Telecommunications, The University of New South Wales, Sydney, New South Wales 2052, Australia}
\author{Thaddeus D. Ladd}
\affiliation{HRL Laboratories, LLC, 3011 Malibu Canyon Road, Malibu, CA 90265, USA}

\begin{abstract}
We design and analyze a logical qubit composed of a linear array of electron spins in semiconductor quantum dots.  To avoid the difficulty of fully controlling a two-dimensional array of dots, we adapt spin control and error correction to a one-dimensional line of silicon quantum dots.   Control speed and efficiency are maintained via a scheme in which electron spin states are controlled globally using broadband microwave pulses for magnetic resonance while two-qubit gates are provided by local electrical control of the exchange interaction between neighboring dots.  Error correction with two-, three-, and four-qubit codes is adapted to a linear chain of qubits with nearest-neighbor gates.  We estimate an error correction threshold of $10^{-4}$.  Furthermore, we describe a sequence of experiments to validate the methods on near-term devices starting from four coupled dots.
\end{abstract}

\maketitle

\section{Introduction}
Proposals for quantum-computing hardware and quantum error correction present compelling visions for quantum information processors~\cite{Loss1998,Kane1998,Imamoglu1999,Kielpinski2002,Taylor2005,Knill2005,Mariantoni2011,Fowler2012,Monroe2013,Devoret2013,Awschalom2013,Gambetta2015,Terhal2015}.  State-of-the-art experiments now involve operations on two to nine coherently controllable qubits~\cite{Lanyon2011,Schindler2011,Zhang2012,Martin2012,Reed2012,Lucero2012,Shulman2012,Yao2012,Cai2013,Nigg2014,Waldherr2014,Chow2014,Cramer2015,Veldhorst2015,Riste2015,Corcoles2015,Kelly2015}, but an extensible logical qubit has not yet been demonstrated.  This paper proposes an experimentally realizable logical qubit in quantum dots using recently demonstrated control of single-electron spins.  Rather than focusing on the scaling issues for a full-scale quantum processor, we instead study in detail how a single logical qubit could work with the limitations of a quantum-dot device having nearest-neighbour gates in a linear array~\cite{Zajac2016}.  The proposal culminates in an ``experimental path'' of demonstrations that build in complexity and reach a quantum-dot logical qubit.

This logical-qubit proposal mirrors work in other quantum-information platforms.  Experiments with multiple coupled qubits have demonstrated proof-of-principle computations and preliminary steps towards a logical qubit.  The field of quantum-processing technology includes photons~\cite{Martin2012,Cai2013}, trapped ions~\cite{Lanyon2011,Schindler2011,Nigg2014}, superconducting qubits~\cite{Reed2012,Lucero2012,Chow2014,Riste2015,Corcoles2015,Kelly2015}, and spins in diamond~\cite{Waldherr2014,Cramer2015}, gallium arsenide~\cite{Petta2005,Nowack2011,Shulman2012,Medford2013,Delbecq2014,Martins2016,Malinowski2016,Nichol2016}, and silicon~\cite{Maune2012,Veldhorst2014,Weber2014,Kawakami2014,Muhonen2014,Dehollain2014,Veldhorst2015,Eng2015}.  However, there are unique advantages to a silicon quantum processor, and the potential for high-fidelity control of long-lived spin qubits motivates this proposal.  We specifically focus here on quantum dots formed in silicon metal-oxide-semiconductor (SiMOS) structures, but also address issues relevant to an implementation of the scheme using quantum dots formed in silicon/silicon-germanium heterostructures.

Single electron spins in isotopically enriched silicon can have coherence times much greater than a millisecond~\cite{Tyryshkin2012,Veldhorst2014,Muhonen2014,Eng2015}, and electrically controlled exchange gates can be performed in tens of nanoseconds~\cite{Petta2005,Nowack2011,Maune2012,Medford2013,Veldhorst2015,Eng2015,Reed2016,Martins2016}.  Electron spins can also be controlled using microwave magnetic resonance, for which high-fidelity gates have been demonstrated~\cite{Veldhorst2014,Fogarty2015,Kawakami2016}.  Quantum dots have a promising path for extensibility since they are compatible with techniques for semiconductor fabrication and integration that were developed for classical computing, though the small feature sizes pose near-term challenges.  A linear array of exchange-coupled dots is perhaps the most accessible design in which to demonstrate a logical qubit.

The proposed quantum-dot logical qubit is within reach of near-term experiments, but it also has the potential for extending to multiple logical qubits.  The hardware platform is a linear array of silicon quantum dots where control operations are restricted to the electrically controlled exchange interaction between neighboring dots and global magnetic-resonance pulses that target all electron spins.  The simplicity of the control scheme is favorable for producing multiple-dot devices, and we show how to adapt simple error correction such as repetition codes to this hardware.  The components of the error correction scheme can be demonstrated in intermediate proof-of-concept experiments, as has been done in other qubit technologies~\cite{Schindler2011,Reed2012,Yao2012,Nigg2014,Waldherr2014,Cramer2015,Riste2015,Corcoles2015,Kelly2015}.  

We assert that a logical qubit must have four characteristics to be extensible, in addition to the DiVincenzo criteria for a quantum computer~\cite{DiVincenzo1997}.  These are:
\begin{enumerate}
\item \emph{Error threshold} --- The system must be able to run an error-detection procedure that itself introduces errors at some tolerably low probability to allow for a threshold error rate~\cite{Aharonov1997,Steane2003}.  For stabilizer codes, this error detection is stabilizer measurement~\cite{Shor1996,Gottesman1998b,Preskill1998}.
\item \emph{Fault tolerance} --- Any single fault is detectable, meaning that the logical qubit must be able to detect errors on its constituent physical qubits in all single-qubit Pauli bases~\cite{Shor1995}.
\item \emph{Parallel measurements} --- The logical qubit must have the ability to perform error-detection measurements at multiple locations simultaneously, where an extensible system has a number of measurement apparatuses proportional to the number of physical qubits~\cite{Steane1998}.  Otherwise, error detection will not keep pace with error generation as the system extends.
\item \emph{Extensible encoding} --- The logical qubit must have an encoding strategy that is capable of extending to correct any number of errors~\cite{Aharonov1997,Steane1998,Preskill1998,Gottesman1998,Nielsen2000,Steane2003}.  For stabilizer codes, this means code distance can increase without compromising any of the preceding criteria.
\end{enumerate}
The logical qubit proposed here is designed to satisfy all of these criteria.  Though the requirements might seem obvious, the sequence of experiments in Section~\ref{sec::experiments} is designed to specifically demonstrate each of these capabilities in silicon quantum dots.

The scope of this paper is a proposal to design and test the simplest logical qubit in a linear array of silicon quantum dots.  Based on recent results in SiMOS dots~\cite{Khoi2013,Veldhorst2014,Veldhorst2015}, we design spin-control protocols and error-correction instruction sequences.  The hardware instructions and error correction are co-adapted to each other, as device fabrication favors simplicity while error correction favors more control of the qubits.  Finding a viable experiment path to satisfy these competing design challenges is the central result of this paper.  Our error correction schemes are simple two-, three-, and four-qubit quantum codes that have been mapped to the linear array of qubits~\cite{Fowler2004,Stephens2008,Stephens2009}, because alternatives like the surface code~\cite{Raussendorf2007,Bombin2007b,Fowler2012,Stephens2014} and Bacon-Shor code~\cite{Bacon2006,Aliferis2007,Cross2009} are not effective in a linear geometry~\cite{Bravyi2009,Bravyi2013,Pastawski2015}.  Our logical qubit is supported with simulations of error correction that can be compared with other proposals~\cite{Preskill1998,Aliferis2006,Szkopek2006,Svore2007,Raussendorf2007,Aliferis2007,Stephens2008,Stephens2009,Cross2009,Fowler2009,Wang2010,Duclos2010a,Wootton2012,Stephens2014,Stephens2014b,Terhal2015}.  Finally, we are careful to note that a purely linear architecture is not extensible to an arbitrary number of qubits, for the simple reason that a single defective qubit anywhere results in two non-interacting arrays.  Our present scope is limited to a logical qubit requiring at most 20 dots, so we do not examine this matter in detail.  However, to establish viability of the proposal, we comment briefly in Section~\ref{sec::discussion} on strategies for handling imperfect dot yield using results from faulty quantum-communication networks.

This paper is structured to show how the capabilities of the quantum-dot hardware and the instruction scheduling for error correction are closely integrated.  The control operations in Section~\ref{sec::control} are designed to be minimal, supporting extensibility, yet sufficient for the error correction in Section~\ref{sec::QEC}.  The proposed quantum-dot platform limits the set of control instructions to favor simplicity in the hardware, but the error correction must adapt to this restrictive control.  The building blocks of error correction in Section~\ref{sec::QEC} form a natural sequence of experiments, described in Section~\ref{sec::experiments}, for culminating in a logical qubit.  The information gained from each experiment is directly related to the role of the QEC building blocks in the ultimate logical qubit.  This experimental path provides milestones towards a logical qubit, and the measured performance of the building blocks can be used to predict performance of a logical qubit.

\section{Controlling Spins in a Linear Array}
\label{sec::control}
\begin{figure}
	\centering
  \includegraphics[width=\columnwidth]{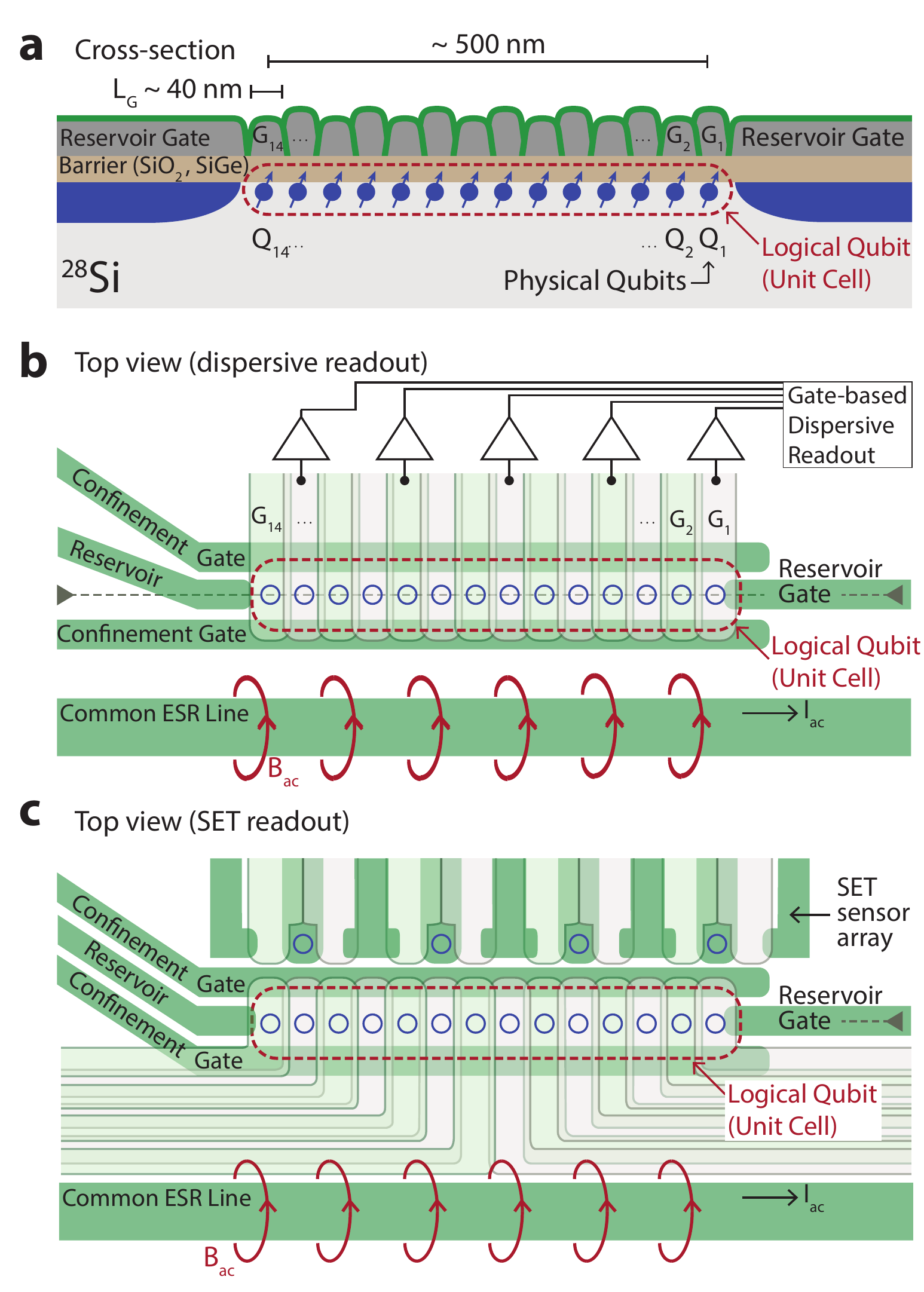}
  \caption{Device schematic for a linear array of quantum dots.  (a)~Cross-section view of the device stack, with dots forming under metal gates.  (b)~Top view of the device where dispersive readout is implemented through the gate electrodes~\cite{Colless2013,Schmidt2014,Stehlik2015}.  (c)~Alternative top view where readout is implemented using single-electron transistors (SETs) located near the dots~\cite{Veldhorst2014,Veldhorst2015}.  This example has 14 dots, which corresponds to a logical-qubit demonstration described in Section~\ref{sec::experiments}.}
  \label{fig::example_device}
\end{figure}

This section describes the hardware platform for the proposed logical qubit, with an emphasis on reducing device complexity as much as possible while still supporting error correction.  Figure~\ref{fig::example_device} depicts a device architecture for a line of exchange-coupled quantum dots, similar to the devices demonstrated in Refs.~\onlinecite{Veldhorst2015,Zajac2016} and employing a microwave ESR antenna as in Ref.~\onlinecite{Dehollain2013}.  In this section, we first present the chosen methods for spin initialization, control, and measurement supported by this architecture.  Second, we perform numerical simulations to estimate performance and identify areas of emphasis in characterizing and mitigating noise.  Finally, we show how the sequencing of control operations, which we call ``tick-tock control,'' implements an instruction set that is sufficient for quantum error correction.  This transition to a logical-qubit encoding is specifically adapted to this SiMOS proposal to work around limitations in the available spin-control operations.

\subsection{Fundamental Control Operations}
The set of spin-control operations is designed to be as simple as possible to support a logical qubit.  Furthermore, all control operations needed for the logical qubit have been demonstrated in silicon quantum dots.  These include electron-spin resonance (ESR)~\cite{Tyryshkin2012,Veldhorst2014,Kawakami2014,Kawakami2016}, electrically controllable exchange interaction between spins in neighboring dots~\cite{Petta2005,Nowack2011,Maune2012,Medford2013,Eng2015,Veldhorst2015}, and electrically controllable preparation and measurement of two spins in the singlet/triplet basis using Pauli blockade~\cite{Johnson2005,Shaji2008,Liu2008,Maune2012,Eng2015}.

To favor extensibility, the logical qubit implements global ESR addressing of all spins (as in an ensemble experiment)~\cite{Tyryshkin2012,Pica2015,Laucht2015}, instead of selective ESR addressing of single spins~\cite{Kawakami2014,Veldhorst2014,Kawakami2016}.  Single-spin addressing is in principle possible with narrow-band microwave pulses, but this approach becomes increasingly difficult with many spins in the device, due to the increasing number of distinguishable frequencies which must be handled within a limited control bandwidth (a problem known as ``frequency crowding"~\cite{Schutjens2013,Riste2015}).  Addressing all spins simultaneously with broadband ESR avoids frequency crowding and the associated cross-talk errors, though this approach clearly limits what control is possible.

Global ESR is also beneficial for dynamical decoupling~\cite{Viola1999}.  Dynamical decoupling is needed for two reasons in our proposal.  First, the combination of an external applied magnetic field and an inevitable distribution in electron $g$-factors leads to an inhomogeneous distribution of Zeeman energies. By applying echo pulses simultaneously to all spins, as practiced routinely in bulk magnetic resonance in inhomogeneous magnetic fields~\cite{Tyryshkin2012,Sigillito2014}, one can coherently control many spins with broadband pulses without requiring a reference oscillator for each spin.  (We note, however, that when the number of spins is low enough, the capability of selective ESR addressing of single spins is useful for testing and calibration.)

A second reason for dynamical decoupling is to correct for dynamic phases that occur during two-qubit operations in our proposal, which employ electrically gated exchange interactions in the presence of large and controllable $g$-factor differences.  As we discuss in the next section and has been demonstrated experimentally~\cite{Veldhorst2015}, these combined phenomena enable two-qubit controlled-phase (CZ) or controlled-NOT (CNOT) gates.  As elaborated in Ref.~\onlinecite{Witzel2015} in the context of donors in silicons, these gates employ some robustness to high-frequency noise due to their use of adiabatic modulation of energy gaps, but they do incur dynamic phases during the adiabatic ramping.  While such phases could be tracked in principle, they are also subject to low-frequency noise sources, and so it is preferred to refocus these phases entirely using dynamical decoupling.

The device schematic shown in Fig.~\ref{fig::example_device} is configured with one gate electrode per quantum dot.  The insulating oxide (such as AlO$_x$) between the metal gates produces a natural tunnel barrier between adjacent dots, as has been observed in experiments on SiMOS two-qubit devices~\cite{Veldhorst2015}.  The exchange coupling between adjacent qubits $Q_i$ and $Q_{i+1}$ is achieved by applying a differential voltage between gates $G_i$ and $G_{i+1}$, to ``detune'' the electric potentials of the two dots.  It is also possible to configure devices with an additional ``exchange'' gate between each pair of qubit control gates (labeled $G_i$ here), and such an approach has been used in Si/SiGe quantum dot devices~\cite{Eng2015,Zajac2015,Reed2016,Zajac2016}. If no exchange gate is used, then it will be necessary to adjust all gates simultaneously, via a self-consistent algorithm, to correct
for the effect of cross-talk between gates when
an exchange operation is applied between a specific pair
of qubits, or pairs of qubits.  Without electrical control of the tunnel coupling, the exchange energy $J$ between each pair of spins is controlled by detuning their relative electrochemical potential; turning on $J$ for one pair will require similarly shifting the electrochemical potentials of dots to the left and to the right to prevent unintended exchange with neighboring spins.  A potential solution to applying exchange between any non-overlapping spin pairs simultaneously is to set the potential at each dot to one of two values, $V^{(0)}$ and $V^{(1)}$.  Starting from one end of the line of dots, assign potentials $(0)$ or $(1)$ such that neighboring dots have the same potential to turn off exchange and different potentials to ``detune'' and activate exchange.  Since $J$ is an exponential function of voltage, inhomogeneity in tunnel coupling can be handled by shifting voltages as needed only slightly from this simplistic model, so that small detunings will still have negligible exchange while large detunings will be designed to match the tunnel coupling of any dot pair and implement a uniform exchange operation in constant time for all dots.

In addition to global ESR for single-qubit control and dynamical decoupling, as well as the exchange interactions for two-qubit gates, our fundamental control operations include singlet preparation for ancilla qubits.  A spin singlet $\ket{S} = (\ket{\uparrow\downarrow}-\ket{\downarrow\uparrow})/\sqrt{2}$ is simple to prepare in a small dot since it is the two-electron ground state.  When a second-electron tunnels into a quantum dot from a thermal bath, a rapid equilibration generates the singlet ground state as long as the electron temperature and the electron Zeeman energy are substantially less than orbital or valley energies; temperatures around 100 mK and fields on the order of a tesla are easily sufficient for high-fidelity singlet preparation in SiMOS, where singlet-triplet splittings often exceed 1 meV~\cite{Lai2011}.   This process may also be reversed for projective singlet-triplet measurement; if a double quantum dot is electrostatically biased into a regime where a two-electron state in a single dot is the ground state, the singlet will occupy this ground state while any symmetric spin-triplet will occupy an excited state due to Pauli blockade~\cite{Ono2002}.  The distinguishable charge signature of the excited state enables distinguishing spin singlet from triplet via charge sensing.  See Refs.~\onlinecite{Petta2005,Lai2011,Maune2012,Eng2015} for more discussion.  The next section examines the performance of these control operations, and Section~\ref{sec::tick_tock} describes how to implement all of the gates needed for error correction.

\subsection{Experimental State of the Art and Simulated Performance in SiMOS Qubits}
All of the spin-control techniques in preceding sections have been demonstrated experimentally in silicon quantum dots.  To support a logical qubit, important measures of performance for each operation are speed and fidelity, in an extensible platform.  Most of the control operations have been rigorously benchmarked, and here the results are already approaching the low error rates required for a logical qubit: high-fidelity singlet preparation, measurement in the singlet-triplet basis, ESR control of individual spins, and memory lifetimes exceeding a millisecond.  An exchange-based CZ gate was recently demonstrated~\cite{Veldhorst2015}, and as we discuss below this gate could have a fidelity sufficient to support QEC for a reasonable level of charge noise.  This section analyzes the recent experimental demonstrations and applies numerical simulations to predict the performance of control operations in a logical qubit.

The Hamiltonian describing qubit control in this section concerns the spins of two singly-occupied dots, $a$ and $b$, with spin operators $\vec{S}_j$ and total $z$-spin projection $m=m_a+m_b$. This Hamiltonian may be written
\begin{multline}
H(t) = \overline{g}[V(t)]\mu_B B_0(S_a^z+S_b^z)
+ \Delta g[V(t)] \mu_B B_0 \frac{S_a^z-S_b^z}{2} \\
+ J[V(t)] \vec{S}_a\cdot\vec{S}_b + \Omega(t) (S_a^x+S_b^x).
\label{Heq}
\end{multline}
where $\overline{g}[V(t)]$ is the average of and $\Delta g[V(t)]$ the difference of $g$-factors for the two dots; these are both functions of the time-dependent applied detuning voltage $V(t)$.  These $g$-factor differences cause differences in the Zeeman energy between dot pairs, which we notate as $\Delta E_Z=\Delta g[V(t)] \mu_B B_0$.  The exchange interaction energy $J[V(t)]$ is also a function of $V(t)$.  ESR-based spin manipulations are implemented via transverse microwave magnetic fields with modulated Rabi frequency $\Omega(t)/2$.  Note that external field $B_0$ points in the $\hat{z}$ direction and the oscillating field $\Omega(t)$ points in the $\hat{x}$ direction.

Concerning non-coherent behavior, it has long been expected, as well as observed in ensemble studies, that electron spins in enriched silicon have long coherence times~\cite{Gordon1958, Kane1998,Vrijen2000,Wolf2001,Tyryshkin2003, Tyryshkin2012}.  Recent experimental demonstrations on single-spin qubits have validated this expectation.  The relaxation time $T_1$ is greater than 1 second~\cite{Yang2013}, so coherence time is limited by spin dephasing due to fluctuations in the magnetic environment.  With a concentration of 800 ppm $^{29}$Si, a dephasing time $T_2^*$ as long as 120 $\mu$s has been demonstrated~\cite{Veldhorst2014}.  Recent investigations have examined the extent to which $T_2^*$ is limited by low-frequency magnetic noise~\cite{Muhonen2014,Fogarty2015}, which can be suppressed with dynamical decoupling schemes.  The coherence time with decoupling has been extended to 1.2 ms with one pulse and 28 ms with multiple pulses~\cite{Veldhorst2014,Veldhorst2015}.  A donor-bound spin in enriched silicon achieved an even longer coherence time~\cite{Muhonen2014}, showing that there is further opportunity for improvement.  Spin-control operations have been experimentally implemented in 1 $\mu$s or less~\cite{Veldhorst2014,Veldhorst2015}, which is four orders of magnitude shorter than the decoupled coherence time.  In the remainder of this section, we will not include these sources of dephasing, and rather focus on control errors, as these are likely to dominate performance in silicon logical qubits.

Preparation and readout in the singlet-triplet basis may utilize spin-to-charge conversion via Pauli blockade~\cite{Taylor2005}.  The method has recently been demonstrated in enriched Si/SiGe devices~\cite{Eng2015} with a readout visibility of 98\%, where loss of visibility includes both preparation and measurement errors.  This experiment was performed at magnetic fields near zero.  At higher magnetic fields, two effects alter the use of Pauli blockade for initialization.  First, at substantial magnetic fields, the $g$-factor variations lead to substantial differences in the Zeeman energy, $\Delta E_Z$, between a dot-pair.  Considering Eq.~(\ref{Heq}), as one ramps from the spin-singlet ground state $(\ket{\uparrow\downarrow}-\ket{\downarrow\uparrow})/\sqrt{2}$ at high $J$ to a lower value of $J$, eventually one reaches a regime where $\Delta E_Z > J$, at which point singlet and $m=0$ triplet states begin to coherently mix.  If intending a fully singlet initialization, one must either ramp quickly enough to avoid this mixing, refocus this mixing with a calibrated pulse sequence, or choose to instead prepare the $\ket{\uparrow\downarrow}$ state by adiabatic ramp down in $J$ as in \cite{Petta2005,Maune2012}.  The latter choice is likely the most robust, and was recently considered for a larger-scale architecture in Ref. \onlinecite{Veldhorst2016b}.  An increased magnetic field may also reduce singlet (or $\ket{\uparrow\downarrow}$) fidelity when the energy splitting between the $m=0$ states and the excited $m=1$ state (i.e. $\ket{T_{-}} = \ket{\downarrow \downarrow}$) decreases with $B$ field.  As outlined below, the operation of the SiMOS device at higher field strengths (up to 1.5T) is beneficial for faster CZ operation times.  The large valley splitting in the SiMOS devices, measured to be $0.3$--$0.8$ meV~\cite{Yang2013} and substantially larger than observed values in Si/SiGe devices~\cite{Eng2015}, permits the use of higher fields before degradation of the Pauli blockade process.  For example, with valley splittings this large, fields of order 1.5 T, and electron temperature around 100 mK, the probability of initializing into a thermal excited state is less than $10^{-6}$ in principle.  In practice, initialization and readout fidelity are limited by noise in control and readout electronics.

Controlled-phase entangling gates based on exchange have been analyzed and experimentally demonstrated extensively in the literature~\cite{Taylor2007, Meunier2011,Veldhorst2015}. The example realized in SiMOS~\cite{Veldhorst2015} is performed via an adiabatic pulse on the electrostatic detuning towards the (0,2) charge-state anticrossing.  If $J(V)$ is the dominant term of this Hamiltonian, the exchange operation would implement swap rotations.  However, the combination of a non-zero $B$ field and a $g$-factor difference $\Delta g(V) = g_a(V) - g_b(V)$ splits the energy levels of the spin states $\ket{\uparrow_a\downarrow_b}$ and $\ket{\downarrow_a\uparrow_b}$, and the nonlinearity in the eigenvalue spectrum near the avoided crossing introduced by $J(V)$ allows a controllable phase shift which has produced controlled-Z and CNOT two-qubit gates between SiMOS quantum dots~\cite{Veldhorst2015}.

\begin{figure}
	\centering
  \includegraphics[width=8.3cm]{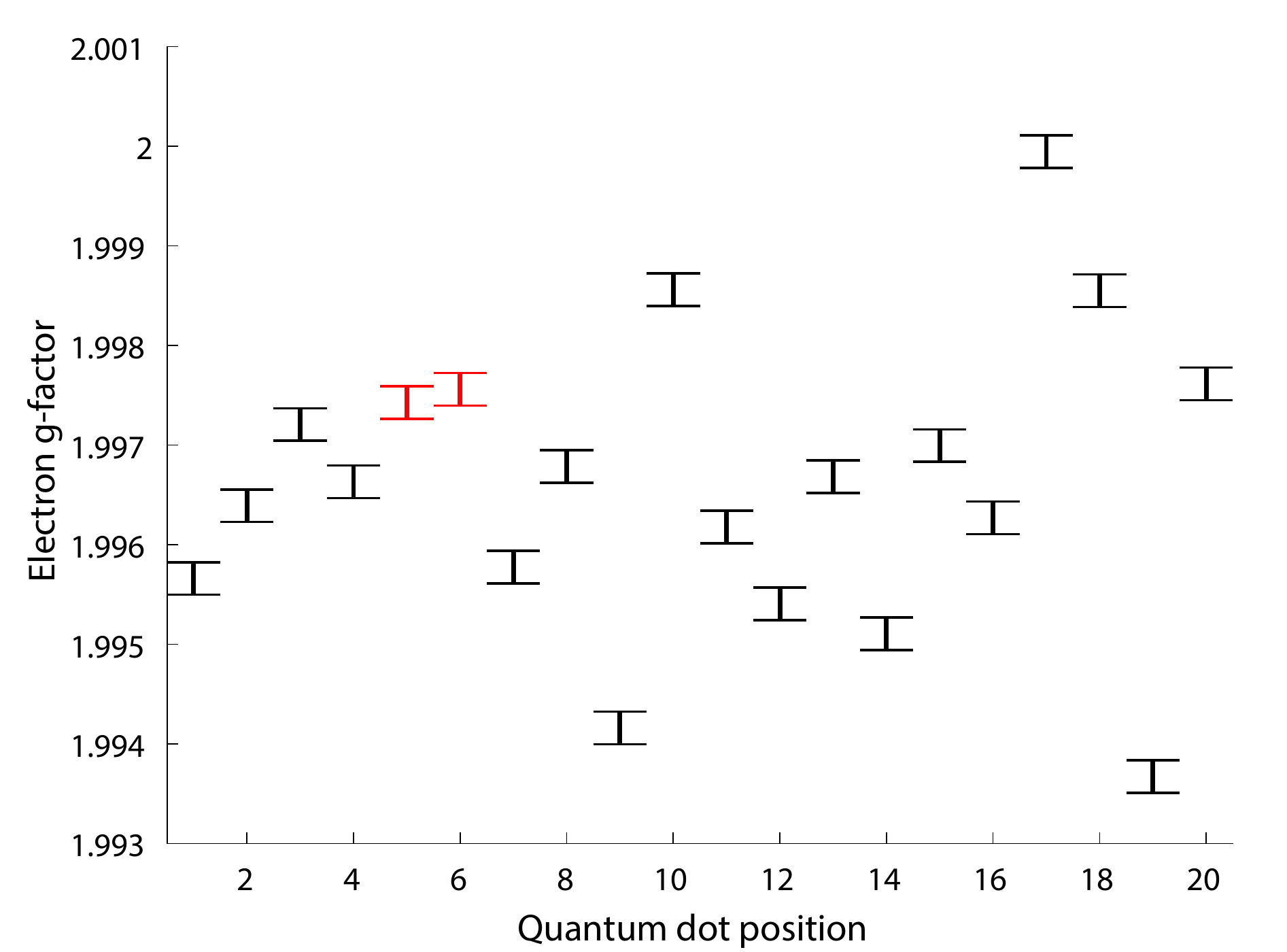}
  \caption{Randomly generated sample of spread in $g$ factors for a linear chain of 20 quantum dots.  The underlying distribution of $g$ factors is based on the measured variance in $g$ factors observed in devices like that in Ref.~\onlinecite{Veldhorst2015}.  Each data point shows the range of $g$-factor tuning possible with Stark shifting~\cite{Yang2013,Veldhorst2014,Veldhorst2015}, corresponding to 10 MHz at $B$ = 1.5~T.  Dots 5 and 6 are colored red to indicate that they have small $\Delta g$ splitting and require the $g$ factors for those dots to be tuned apart for the CZ gate, as described in the text.}
  \label{fig::gfactor_spread}
\end{figure}

As the duration and fidelity of the CZ gate depends in part on $g$-factor differences between dots, it is important to characterize what values of $\Delta g$ are achievable.  Disorder perturbations at the Si/SiO$_2$ interface lead to a stochastic and bias-dependent variation in $g$-factors.  Figure~\ref{fig::gfactor_spread} illustrates a randomly generated distribution of electron $g$ factors for a linear chain of 20 qubits, as well as the $g$-factor tuning range, based on statistics from measurements on SiMOS devices~\cite{Veldhorst2015}.
In the event where two neighboring dots have small $\Delta g$ at zero electrostatic detuning, the difference can be increased with Stark shifting by choosing whether to detune the dots towards (2,0) or (0,2) charge configuration~\cite{Yang2013,Veldhorst2014,Veldhorst2015}, noting that this would yield a favorable configuration for detuning potentials $(0)$ and $(1)$ along the chain as discussed in the previous section.  In a recent experiment, the minimum energy splitting at $B_0 = 1.4$~T was $\Delta E_Z = \Delta g\mu_B B_0= 20$~MHz$\times h$, with $10$~MHz tuneability in each electron spin.

\begin{figure}
	\centering
  \includegraphics[width=6cm]{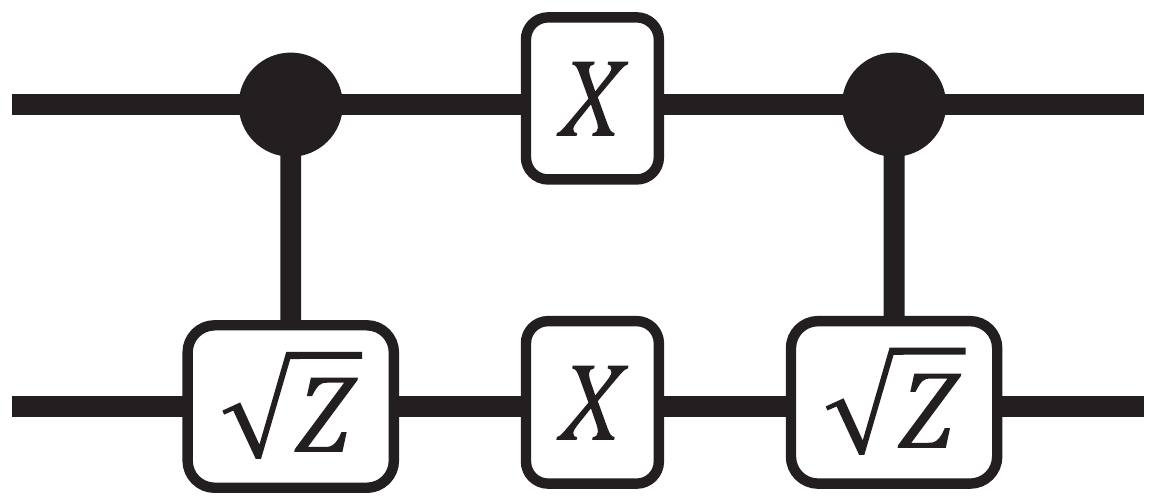}
  \caption{Quantum circuit illustration of the dynamical decoupling scheme for the adiabatic controlled-phase gate; the controlled-$\sqrt{Z}$ pulses on two qubits in two dots are implemented via adiabatic pulsing of exchange.  The resulting operation includes single-qubit $Z$ rotations; these are decoupled by the intervening single-qubit $\pi$ pulses about $X$, which may implemented via a global ESR pulse.}
  \label{fig::CZDD}
\end{figure}

From preliminary estimates, simple square pulsing of the CZ operation with observed values of $\Delta E_Z$ can achieve a two-qubit gate fidelity above 99\%, but substantially higher fidelity is accessible through pulse shaping.  In particular, adiabatic pulsing~\cite{Witzel2015} has several advantages, principle among them being a resilience to some noise processes. In the adiabatic limit, pulsing into the avoided crossing and back realizes a combination of Zeeman phase shifts and a non-linear phase shift due to $J(V)$.  The nonlinear phase shift is given entirely by the time integral of $J(V)$, which we here notate as $\xi=\int J[V(t)]dt/\hbar$.  The integral is over a sufficient time to fully capture a voltage pulse $V(t)$ which brings $J(V)$ to and from a negligibly small value.  The total adiabatic unitary evolution for the two spins is then given by
\begin{multline}
U(\xi) = \\
\begin{aligned}
\exp\biggl\{-i\xi S_a^z S_b^z-i\int&\biggl[\omega_0[V(t)]+\frac{1}{2}\Omega[V(t)]\biggr]dt S_a^z
\\
                                   -i\int&\biggl[\omega_0[V(t)]-\frac{1}{2}\Omega[V(t)]\biggr]dt S_b^z
                                   \biggr\},
\end{aligned}
\end{multline}
where $\hbar\omega_0(V)=\bar{g}\mu_B B^z$ and
\begin{equation}
\hbar\Omega(V) = \sqrt{\Delta E_Z^2(V)+J^2(V)}.
\end{equation}
In practice, the adiabatic limit is maintained by assuring that the frequency bandwidth of a $J[V(t)]$ pulse lies well beneath the minimum value of $\Delta E_Z[V(t)]/\hbar$.  If the total nonlinear phase shift satisfies $\xi = \pi$, one achieves a maximally entangled CZ gate in addition to local single-spin phase shifts.  These single-spin phase shifts are substantial, however, and subject to magnetic and charge noise, the latter due to the electric-field dependence of $g_j$.  Rather than attempting to compensate for these phases and accept errors due to low-frequency magnetic or charge noise, we instead employ an approach where we decouple these phases, as in Ref.~\onlinecite{Witzel2015} and illustrated in Fig.~\ref{fig::CZDD}.  Denote by $X_j$ a $\pi$ pulse for spin $j$, and break our $J(V)$ pulse into two halves each satisfying $\xi = \pi/2$.  Then under perfectly adiabatic conditions,
\begin{equation}
U(\pi/2) X_1 X_2 U(\pi/2)  = e^{-i\pi/2} S_1 S_2 U_{\text{CZ}} X_1 X_2,
\label{eq::CZDD}
\end{equation}
where $U_{\text{CZ}}$ is a controlled-phase gate and $S_j = \sqrt{Z_j}$ is a single-qubit $S$-gate, which may be corrected via $g$-factor manipulation or by a frame update.  No extraneous magnetic phases need be tracked in this decoupled CZ gate.

\begin{figure}[ht!]
	\centering
  \includegraphics[width=8.3cm]{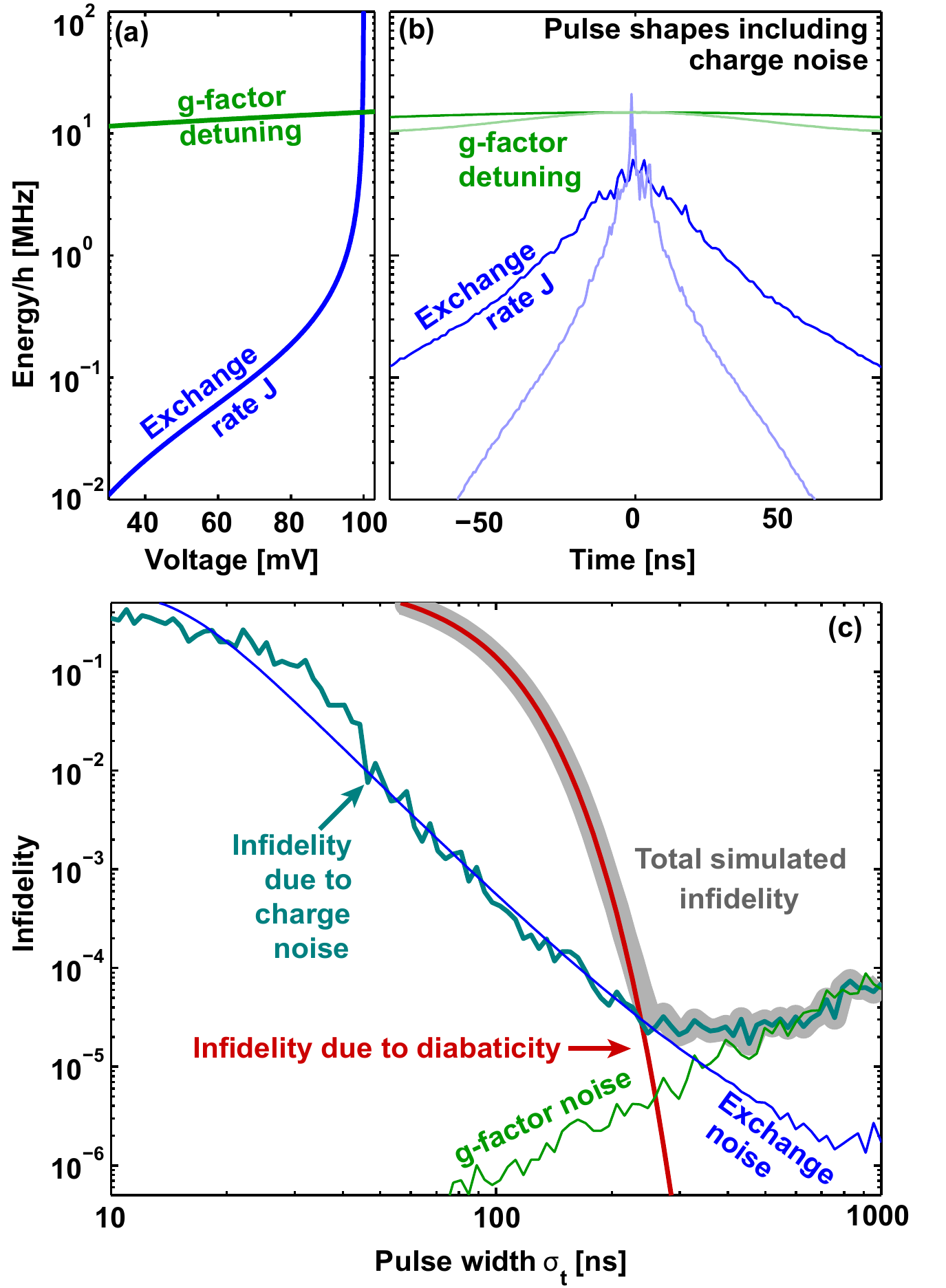}
  \caption{Implementing a controlled-phase gate using exchange and $g$-factor differences.  (a)~The modeled functions of $g$-factor difference $\Delta g(V)$ exchange rate $J(V)$ versus detuning voltage, employing typical parameters, similar to those demonstrated in Ref.~\onlinecite{Veldhorst2015}.   (b)~Pulse shapes for $\Delta g[V(t)]$ and $J[V(t)]$, plotted on the same scale as panel (a), for a Gaussian detuning voltage pulse $V(t)\propto \exp(-t^2/2\sigma_t^2)$ for two different values of $\sigma_t$, including added voltage noise $\delta V(t)$ with spectral noise density $S_V(t) = A^2/f$.  These sample traces use a rather high value of $A\sim 30$~$\mu$V to allow the noise to be visible on this scale.  (c)~Simulated infidelity of the adiabatic CZ gate.  The red line results from simulating with no charge noise but examining infidelity due to non-adiabatic behavior.  The cyan line considers strictly adiabatic evolution but adds charge noise by Monte-Carlo integration using sampled voltage noise as in (a), in this example with $A=5$~$\mu$V.  The charge-noise-induced infidelity (cyan line) decreases due to exchange noise, for which the primary trend indicated by the blue line follows $10(A/I_{\text{peak}})^2$, where $I_{\text{peak}} = J/|dJ/dV|$ is the insensitivity~\cite{Reed2016} at peak $J$.  For longer pulses the charge-noise-induced infidelity increases due to noise on the $g$-factor; the green trend plot is $\int |\Delta g[V(t)]|^2dt/(500\text{~Grad/sec})$.  The thick gray line is the total infidelity, estimated as the sum of the diabaticity (red) and charge-noise (cyan) contributions.}
  \label{fig::CZ_gate}
\end{figure}

Two sources of error are expected to limit the fidelity of this CZ gate, and these are considered in simulations indicated by Fig.~\ref{fig::CZ_gate}.  These simulations use a standard detuning model for $J(V)$ and a linear model for $\Delta g(V)$, both informed by Ref.~\onlinecite{Veldhorst2015} and indicated in Fig.~\ref{fig::CZ_gate}(a).  In the $J(V)$ model, the tunnel coupling $t_c$ between dots is assumed to decrease for very small $V$, but saturates to a constant at large detuning $V$, where $J(V)\sim t_c^2/(V_0-V)$~\cite{Maune2012}.  The chosen value and bias dependence of $\Delta g(V)$ for this simulation are typical values from measurements of devices similar to the one in Ref.~\onlinecite{Veldhorst2015}.  Simulations use two Gaussian voltage pulses which satisfy $\xi = \pi/2$ interspersed with ideal single-spin $\pi$ pulses as in Fig.~\ref{fig::CZDD}.  The simulation integrates the evolution due to these Gaussian pulses from $-5\sigma_t$ to $5\sigma_t$, where $\sigma_t$ is the root-mean-square temporal pulse width.

As indicated in Fig.~\ref{fig::CZ_gate}(b), Gaussian pulses in $V(t)$ lead to sharply peaked pulses in $J[V(t)]$ for short $\sigma_t$ and to smoother, broader pulses for long $\sigma_t$.  These shapes are especially critical for the influence of charge noise, which is introduced as a randomly sampled noisy voltage $\delta V(t)$.  Ensembles of $\delta V(t)$ functions are filtered from Gaussian white noise to produce the noise spectral density $S_V(f)=A^2/f$, including a low-frequency cut-off corresponding to a 1-hour calibration timescale.  This $1/f$ voltage noise mimics the expected influence of electric field noise from a variety of possible sources in a real device by modeling it as a single noisy voltage.  A clear noise enhancement at the peak value of $J[V(t)]$ is visible in Fig.~\ref{fig::CZ_gate}(b), especially for the shorter pulse (lighter blue line); this is because the noise insensitivity $I = J/|dJ/dV|$ rapidly decreases at high $J$ for the detuning mode of operation~\cite{Reed2016}.  The result of integrating the Schr\"odinger equation for these chosen pulse shapes is shown in Fig.~\ref{fig::CZ_gate}(c), in which infidelity is given by the normalized trace-distance between the simulated, imperfect unitary and the ideal unitary of Eq.~(\ref{eq::CZDD}) under perfect adiabatic and noise-free conditions.  Figure~\ref{fig::CZ_gate}(c) uses the particular gate-referred charge noise amplitude $A=5$~$\mu$V, a value comparable but somewhat improved relative to observed charge-noise either deduced from CZ oscillation decay in Ref.~\onlinecite{Veldhorst2015} or measured in similar MOS devices in Ref.~\onlinecite{Freeman2016}.

The red curve of Fig.~\ref{fig::CZ_gate}(c) shows the infidelity due to non-adiabatic behavior, which dominates at short pulse widths $\sigma_t$ but then falls rapidly with increasing $\sigma_t$.  The cyan curve indicates infidelity due to randomly sampled $1/f$ charge noise.  This contribution to gate error may be decomposed into two sources.  In the long-pulse limit, the limiting noise comes from charge-noise-induced fluctuations in $g$-factor, since this error increases with pulse length following a trend proportional to $\int |\Delta g[V(t)]|^2 dt$, as indicated by the green line in Fig.~\ref{fig::CZ_gate}(c). At the minimum of infidelity relative to root-mean-square pulse-width $\sigma_t$ the dominant noise source is an imperfect nonlinear phase from the integral over a noisy $J[V(t)]$, which is dominated by charge noise at the peak of the exchange pulse.  This noise source is therefore proportional to $(A/I_{\mathrm{peak}})^2,$ where $I_{\mathrm{peak}}$ is the insensitivity at the peak of the exchange pulse.  This contribution decreases for longer pulses which have a lower peak value of $J$, as indicated by the blue line.  This error source could be reduced with symmetric exchange pulsing if an additional gate were available to modulate the tunnel barrier between dots~\cite{Reed2016,Martins2016}.  There are important tradeoffs to consider between the additional electrostatic tunability offered by exchange gates and the increased device complexity from doubling the number of gates, but we defer the matter to other investigations in the literature~\cite{Veldhorst2015,Reed2016,Martins2016}.    The minimum total infidelity occurs around $\sigma_t\approx 300$~ns for these parameters, at which, over a broad range of $A$, the minimum infidelity scales as $A^2$.  The pulse width providing this minimum of course varies approximately linearly with the constant and voltage-dependent $g$-factor differences, which vary from dot pair to dot pair, but the dependence on these parameters of the minimum fidelity reached at the optimum pulse length is sublinear, allowing a substantial range of $g$-factor variation with infidelity comparable to the simulation shown in Fig.~\ref{fig::CZ_gate}.

Our simulation of CZ infidelity has not included imperfections in single-qubit operations, such as the decoupling $\pi$-pulses.  However, existing experimental implementations of cryogenic ESR give strong encouragement that these pulses can be achieved with high fidelity.  Through the use of an on-chip transmission line~\cite{Dehollain2013}, ESR control of single electron spins in SiMOS devices has been demonstrated with benchmarked control fidelity of as high as 99.6\%~\cite{Veldhorst2014,Fogarty2015}; another experiment with a micromagnet in Si/SiGe dots realized 99\% fidelity~\cite{Kawakami2016}.  However, as the number of electron spins in the device increases, frequency crowding becomes a notable issue, so the global ESR scheme introduced in the previous section would have all spins controlled by a common ESR transmission line.  High-fidelity control of spin ensembles has been demonstrated in magnetic resonance~\cite{Sigillito2014}, where composite pulse sequences like BB1 are used to correct for systematic over-rotation errors~\cite{Wimperis1994,Brown2004}.  Global ESR on all spins in this proposal will require such broadband pulsing since the expected spread in values for electron $g$ factor will lead to a spread in Zeeman energies of tens of MHz at $B_0 = 1$~T.  The ability to electrically Stark shift the electron $g$-factor provides another resource for maintaining high-fidelity control using only global ESR pulses~\cite{Yang2013,Veldhorst2015}.

We finally note that several optimizations of the CZ gate discussed here are available and may be analyzed in future work.  The Gaussian pulse shape for $V(t)$ chosen in our analysis was not an optimized choice; shaped pulse sequences for exchange~\cite{Wang2014} and adiabatic CZ~\cite{Ghosh2013, Barends2014} gates have been employed in other contexts and allow for some optimization of fidelity.  Further, the simple single-pulse dynamical decoupling routine we have employed could be extended to multipulse sequences to further suppress $g$-factor noise, which ultimately limits fidelity for very long pulses.  Optimization of dynamical decoupling, especially in conjunction with the identification of bias regions of high insensitivity, can lead to drastic improvements in fidelity in the presence of charge noise~\cite{Witzel2015}.

\subsection{Tick-Tock Protocol}
\label{sec::tick_tock}
All of the necessary gates for a logical qubit can be produced by deliberate sequencing of the following spin-control operations: preparation of spin singlet, global ESR, and the exchange-driven CZ gate.  We call our scheme for controlling quantum-dot spin qubits ``tick-tock'' control, because control pulses are sequenced into two alternating time intervals, called simply tick and tock.  The transitions between tick and tock are defined by applying global Hadamard gates using ESR~\cite{Tyryshkin2012,Pica2015,Laucht2015}.  Exchange-driven CZ gates are selectively implemented within tick or tock intervals~\cite{Veldhorst2015}.  Finally, two-spin singlets are prepared, coupled to data by CZ gates, and measured in singlet-triplet basis, as explained below.

\begin{figure}
	\centering
  \includegraphics[width=8.3cm]{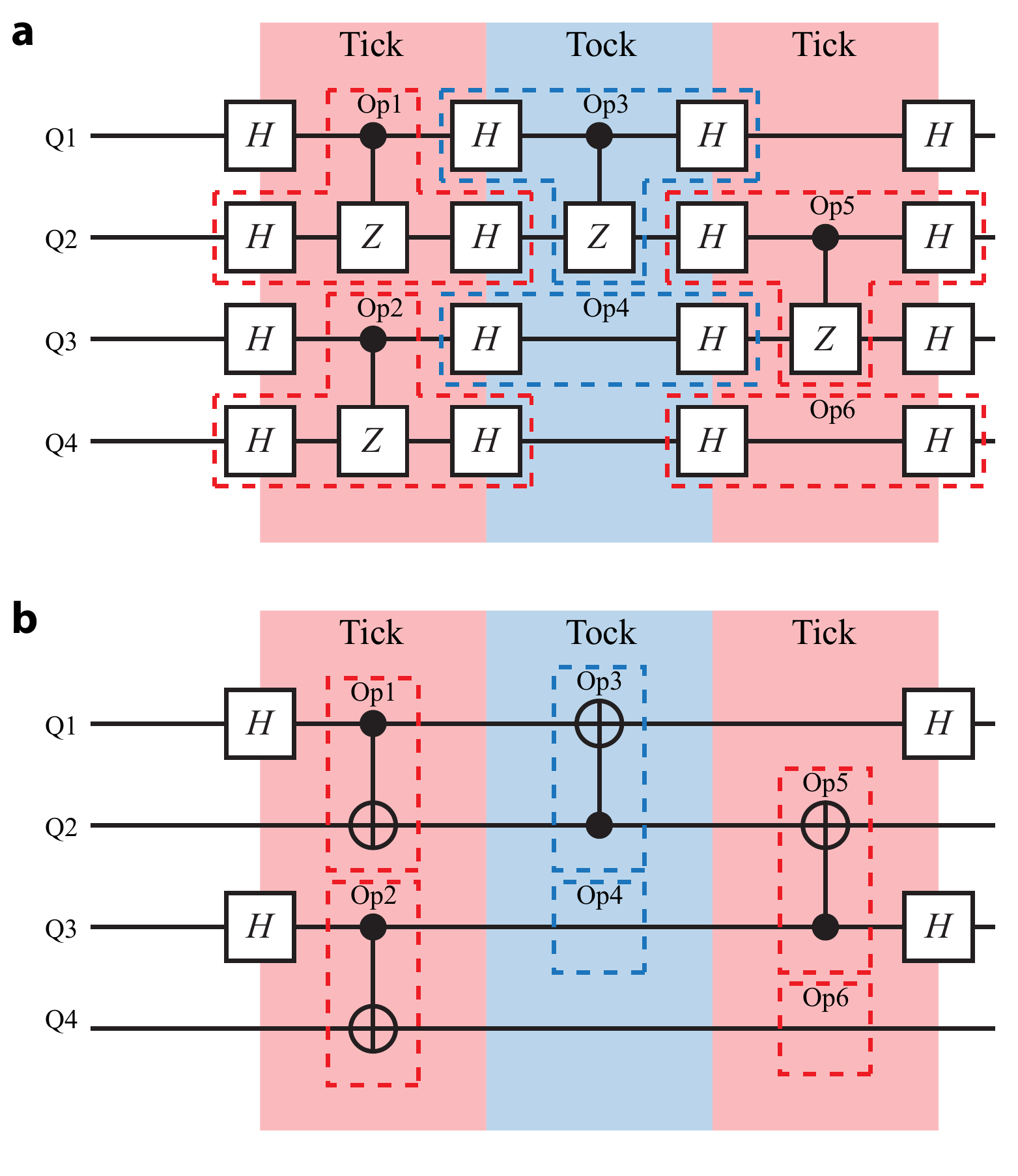}
  \caption{Example of using tick-tock control to apply CNOTs to four spins, labeled Q1--Q4.  (a)~Original control sequence consisting of global Hadamard gates that demarcate transitions between tick and tock intervals, as well as selectively addressed CZ gates within a tick or tock interval.  (b)~Equivalent circuit diagram where two Hadamard gates and a CZ are merged to form a CNOT, with grouping shown by dashed boxes.  The convention is that CNOT will have its control on an odd qubit in tick intervals (Op1, Op2, and Op5) and on an even qubit in tock intervals (Op3).  When there is no intervening CZ gate, Hadamards pair to identity (Op4 and Op6).  As explained in the text, the unmatched Hadamard gates at the beginning and end of computation are not a concern.}
  \label{fig::tick_tock}
\end{figure}

The tick-tock protocol can implement any CNOT between neighboring spins by appropriate timing of the exchange pulse.  The Hadamard gates that transition between tick and tock intervals transform each CZ gate~\cite{Veldhorst2015} into a CNOT gate, using the feature of the Hadamard gate that it interchanges $X$ and $Z$ operators~\cite{Pica2015}.  Figure~\ref{fig::tick_tock} shows a circuit diagram~\cite{Nielsen2000} that illustrates how any CNOT between neighboring qubits can be implemented by selectively performing a CZ gate in the appropriate tick or tock interval and merging with neighboring Hadamard gates.  Unlike CZ, CNOT is not a symmetric gate, so the orientation depends on which qubits participate in the gate and whether it occurs in a tick or tock interval.  The convention here is that the control qubit is odd-numbered in a tick interval and even-numbered in a tock interval.  The unmatched Hadamard gates in Fig.~\ref{fig::tick_tock}b, which only occur at the beginning or end of the experiment, can be ignored since the single-spin data qubits are in an arbitrary state at the beginning and end of the computation.  As explained below, the data spins are initialized using the aid of two-spin ancillas for measurement, after the tick-tock protocol has started.

There are two types of qubits in tick-tock control.  Data qubits that hold logical information will be single spins, and measurement uses two-spin ``ancilla'' qubits which span either the singlet-triplet or ``flip-flop" ($\ket{\uparrow\downarrow}/\ket{\downarrow\uparrow}$) basis.  Within tick-tock control, this ancilla has the additional feature that it can be used in a ``measurement gadget'' to projectively measure a data spin in either $X$ or $Z$ basis, determined by timing of exchange pulses.  Throughout this section, we refer to the ancilla as being in the singlet-triplet basis, although as noted in the previous section, higher initialization fidelity may lead to preferring the ``flip-flop" basis, $\ket{\uparrow\downarrow}/\ket{\downarrow\uparrow}$, which is initialized adiabatically.  An adiabatic initialization procedure can be adapted to the tick-tock protocol by performing the traversal in a tick interval when the CZ coupling gates are to be performed in a tock interval, or vice versa.  
For either basis, when combined with the CNOTs from tick-tock control, we can use this ancilla to make all of the measurements required for error correction.

\begin{figure}
	\centering
  \includegraphics[width=8.3cm]{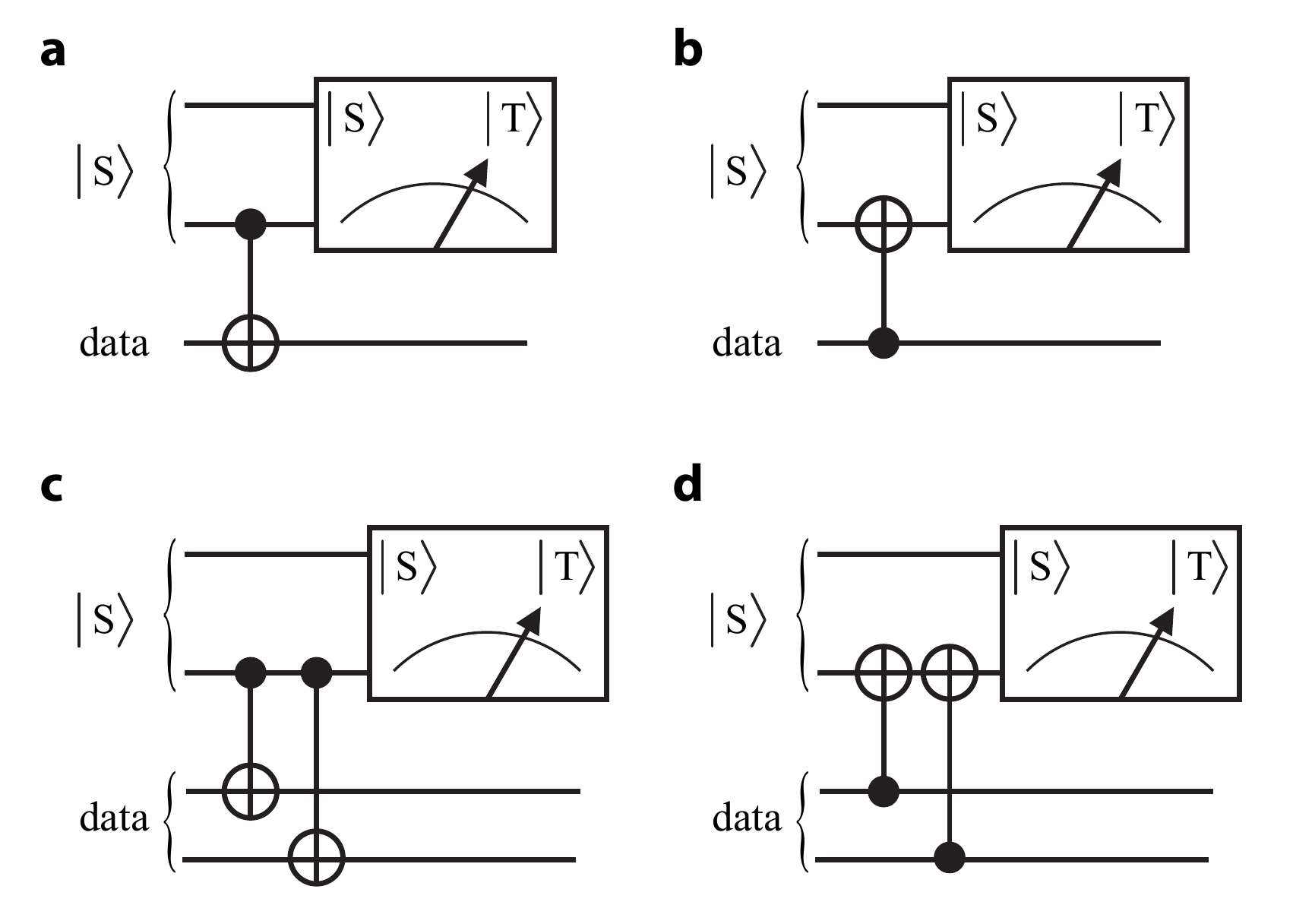}
  \caption{Measurement gadgets using singlet preparation and measurement in singlet/triplet basis: (a)~$X$-basis measurement; (b)~$Z$-basis measurement; (c)~parity measurement of $X \otimes X$ on two data spins; (d)~parity measurement of $Z \otimes Z$ on two data spins.  The gadget extends to measuring a $X$ or $Z$ operator of any size by adding CNOT gates.}
  \label{fig::measure_gadget}
\end{figure}

The measurement gadget is a tool for measuring data spins in $X$ or $Z$ basis.  In addition to measuring a single data spin, the gadget can be extended to projectively measure a multi-qubit $X$- or $Z$-basis operator, such as $X \otimes X$ or $Z \otimes Z$.  Herein, we only consider measuring operators that are purely $X$ or $Z$ type, as this is sufficient for an encoding family known as Calderbank-Shor-Steane (CSS) error correction~\cite{Calderbank1996,Steane1996b,Nielsen2000}.  The measurement gadget works by applying a CNOT to one of the ancilla spins (either works due to symmetry) and a data spin.  To measure in the $X$ basis, use timing in the tick-tock protocol to put the control qubit of the CNOT on the ancilla spin (Fig.~\ref{fig::measure_gadget}a).  Using subscript ``d'' for data and ``a'' for ancilla, the CNOT transformation is
\begin{align}
&\left(\alpha\ket{+}_d + \beta\ket{-}_d\right) \otimes \ket{S}_a \nonumber \\
&\xrightarrow{\text{CNOT}} \alpha\ket{+}_d \otimes \ket{S}_a + \beta\ket{-}_d \otimes Z\ket{S}_a,
\label{eqn::measure_x}
\end{align}
where $\ket{S} = \left(\ket{01}-\ket{10}\right)/\sqrt{2}$ is a singlet, $Z\ket{S} = \left(\ket{01}+\ket{10}\right)/\sqrt{2}$ is one of the triplets (apply Pauli $Z$ to either one of the spins in the singlet), and $\ket{+}$ and $\ket{-}$ are the eigenstates of $X$.  Likewise, to measure in the $Z$ basis, put the target qubit on the ancilla spin (Fig.~\ref{fig::measure_gadget}b).  In this case, the CNOT transformation is
\begin{align}
&\left(\alpha\ket{0}_d + \beta\ket{1}_d\right) \otimes \ket{S}_a \nonumber \\
&\xrightarrow{\text{CNOT}} \alpha\ket{0}_d \otimes \ket{S}_a + \beta\ket{1}_d \otimes X\ket{S}_a,
\label{eqn::measure_z}
\end{align}
where $X\ket{S} = \left(\ket{00}-\ket{11}\right)/\sqrt{2}$ is another triplet (apply Pauli $X$ to either one of the spins in the singlet).  In words, the singlet is converted to triplet if the data spin is $\ket{1}$, the $-1$ eigenstate of $Z$.  For Equations~(\ref{eqn::measure_x}) and~(\ref{eqn::measure_z}), measuring the ancilla as singlet or triplet performs projective measurement in $X$ or $Z$ basis (respectively) on the data spin.

\begin{figure*}
	\centering
  \includegraphics[width=16cm]{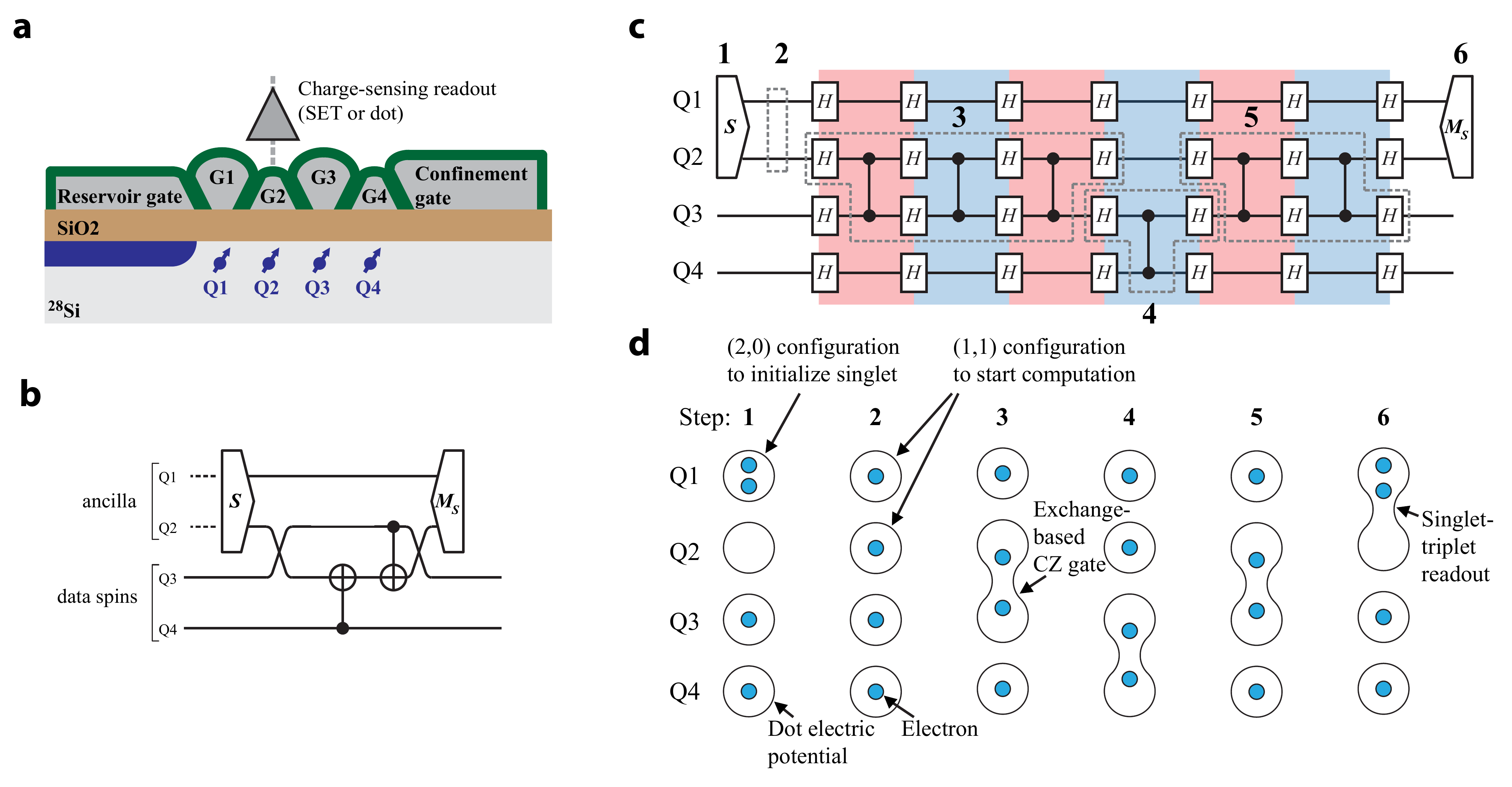}
  \caption{Device schematic and control sequence for parity-measurement experiment.  These diagrams show how an error-correction control sequence is mapped from circuits to operations on quantum dots. (a)~Four exchange-coupled silicon dots with reservoir.  (b)~Parity measurement circuit using a singlet ancilla and measurement.  The two-qubit gates are standard-LNN instructions, described in Section~\ref{sec::QEC}.  This representation is convenient because it is compact, but an experiment must unpack each instruction into a sequence of ESR and exchange operations.  (c)~Tick-tock control sequence showing how the two-qubit gates in panel b are decomposed into CZ entangling gates between the global Hadamards implemented by ESR.  SWAP is decomposed into three CNOTs, and the CNOT-followed-by-SWAP gate is two CNOTs, using a circuit identity~\cite{Nielsen2000}.  (d)~Sequence of electrostatic voltage biasing to implement the exchange-based CZ gates, with step numbers corresponding to panel c and dots labeled Q1--Q4 as in the other panels.  Composite gates such as those shown in steps 3, 4, and 5 of panel c may require multiple exchange pulses, so what is depicted in panel d shows which spins are undergoing exchange.}
  \label{fig::parity_detection}
\end{figure*}

The gadget expands to projectively measuring any multi-qubit $X$ or $Z$ operators by applying a CNOT with orientation specified above between the ancilla and each data spin covered by the operator.  For example, when measuring $Z \otimes Z$ on two data spins, the ancilla will flip between $\ket{S}$ and $X\ket{S}$ for each data spin in the $\ket{1}$ state, and likewise between $\ket{S}$ and $Z\ket{S}$  for $X$-basis measurement.  Measurement of a two-qubit operator $X \otimes X$ (Fig.~\ref{fig::measure_gadget}c) or $Z \otimes Z$ (Fig.~\ref{fig::measure_gadget}d) is the fundamental operation in ``parity-measurement'' experiments that have been demonstrated in other qubit technologies~\cite{Schindler2011,Reed2012,Yao2012,Nigg2014,Waldherr2014,Cramer2015,Riste2015,Corcoles2015,Kelly2015}.  Figure~\ref{fig::parity_detection} shows how to implement the parity-measurement gadget in a device with four quantum dots.  This parity-measurement gadget is the first demonstration in the experimental path described in Section~\ref{sec::experiments}.  Moreover, all of the experiments use the parity measurement gadget as a subroutine in codes for a logical qubit, so they are extensions of the procedure depicted in Fig.~\ref{fig::parity_detection}.

In addition to measuring parity, the measurement gadget is also used to initialize data spins.  The data spins are loaded into dots in an arbitrary mixed state.  Then the tick-tock protocol of periodic Hadamard gates is initiated, and measurement gadgets are used to prepare each data spin in either $X$ or $Z$ basis as needed for the computation.  An extensible logical qubit will require readout apparatuses that are regularly spaced in the array of dots, so this initialization procedure can be performed in constant time.  An alternative initialization technique is to prepare two data spins as a singlet, which is a valid encoded state for some error-correcting codes.  The final operation needed for universality is magic-state injection, which can be achieved by preparing a $\ket{+}$ state using the measurement gadget and then using a voltage pulse on that dot to Stark-shift the electron $g$ factor, implementing a phase rotation.  For example, a $\pi/4$ phase shift prepares the magic state for a $T$~gate, which can be distilled with other available gates, as studied in the literature~\cite{Bravyi2005,Raussendorf2007,Jones2012}.

\section{Logical Qubit in One Dimension}
\label{sec::QEC}
Implementing a logical qubit requires detailed instruction sequences to perform encoding, decoding, and logical gates that incorporate error detection.  The tick-tock scheme in Section~\ref{sec::control} places significant constraints on which gates are available, though it is still sufficient to implement error correction.  In short, the main challenge for the proposed logical qubit is that any encoding scheme requires two-qubit gates between qubits that cannot all be local in a linear arrangement.  The non-local interactions must be mediated by SWAP gates, and we discuss the significant prior work in this area below.  This section describes the instruction sequences for two- and three-qubit repetition codes, as well as a four-qubit error detection code.  By convention, the codes are identified by the number of data qubits, though ancillas for error detection are also required.  These codes are all closely related to the error correction proposal by Shor~\cite{Shor1995}, and they appear to be the simplest codes that can be implemented in a linear array.  The codes also provide a sequence of experiments that demonstrate the essential features of a logical qubit, which are described in Section~\ref{sec::experiments}.

We briefly describe the notation used in this section.  The Pauli operators will be denoted as $X$ or $Z$, and for multi-qubit operators the tensor product will be implicit such as $XX$.  When needed, subscripts will index which qubit a Pauli operator acts on, and missing subscripts are implicitly the identity; for example, $X_1 X_3$ is a tensor product with identity operator on the second qubit and any others in the system.  The +1 eigenstate of the $X$ operator is denoted $\ket{+} = (\ket{0} + \ket{1})/\sqrt{2}$.  Encoded operators and states for a quantum code are denoted with a bar, such as $\bar{X}$ or $\ket{\bar{0}}$.  Diagrams in this section use the quantum-circuit representation for its compactness, and these diagrams can be expanded to implement the tick-tock protocol as shown previously in Fig.~\ref{fig::parity_detection}.  The ``weight'' of a Pauli operator is the number of non-identity terms in its tensor-product expansion into single-qubit Pauli operators~\cite{Nielsen2000}; for example, $\mathrm{weight}(X_1 X_3) = 2$.

\subsection{Background on Error Correction in Constrained Geometries}
For many qubit technologies, including quantum dots, long-range coupling is challenging.  Several investigations into quantum error correction attempted to address this problem by studying codes that require only local interactions for qubits on a lattice in a finite number of dimensions.  The toric code introduced by Kitaev~\cite{Kitaev2003} was specifically designed to have local stabilizer measurements in two dimensions (albeit the surface of a torus).  The surface code and cluster-state computation emerged as variants of the toric code, preserving the important local-stabilizer feature while introducing boundaries for planar embedding or otherwise modifying the code to suit a particular architecture~\cite{Dennis2002,Raussendorf2007,Bombin2007b,Fowler2009,Devitt2009,Fowler2012,Jones2012}.  Another code family with similar properties are the color codes~\cite{Bombin2007,Landahl2011,Stephens2014b,Landahl2014,Kubica2015,Brown2015,Bombin2015,Jones2016,Criger2016}, which also have local stabilizers.  Surface and color codes are prominent examples of topological codes, which are codes that have local stabilizers and increase code distance by extending the size of the code~\cite{Pastawski2015}.  A code with similar properties is the Bacon-Shor code~\cite{Bacon2006,Aliferis2007}, which is a subsystem code with local ``gauge'' operators in two dimensions.  However, it is not topological because its stabilizers are not local.

Topological codes are not suitable for a linear nearest-neighbor (LNN) architecture because they cannot have a threshold in one dimension~\cite{Bravyi2009,Bravyi2013,Pastawski2015}.  Nevertheless, these codes provide instructive lessons.  Many topological codes have good thresholds~\cite{Raussendorf2007,Fowler2009,Wang2010,Duclos2010a,Stephens2014,Stephens2014b,Li2017}, and this seems to result from the local stabilizers~\cite{Stephens2014,Stephens2014b}.  Specifically, local stabilizers can be measured with short sequences of gates, limiting the potential for error propagation.  Although the codes in this proposal are not topological, the stabilizer-measurement circuits are similarly compact; they use one- or two-qubit ancillas for low-weight measurements, as in surface codes~\cite{Dennis2002,Raussendorf2007,Fowler2009,Devitt2009,Duclos2010a,Horsman2012,Fowler2012} and Bacon-Shor codes~\cite{Bacon2006,Aliferis2007,Cross2009}.

As we stated in the Introduction, a logical qubit must have the ability to increase code distance.  The main alternative to topological codes is code concatenation, where codes are nested inside of codes~\cite{Nielsen2000}, which is the approach taken in this proposal.  There have been encouraging results in thresholds with concatenation~\cite{Steane2003,Knill2005,Poulin2006,Cross2009}, as well as investigations into two-dimensional and LNN architectures~\cite{Fowler2004,Szkopek2006,Svore2007,Aliferis2007,Stephens2008,Stephens2009,Spedalieri2009,Li2017}.  Knill demonstrated that small quantum codes (such as the four-qubit code studied here) can be effective when concatenated~\cite{Knill2005}.  Although the thresholds calculated in that proposal are very high (3\% or greater), the model for qubits assumes arbitrary connectivity that cannot be realized with only nearest-neighbor interactions.  Subsequently, Stephens and Evans developed an implementation of the subsystem four-qubit code in a LNN geometry~\cite{Stephens2009}.  Our four-qubit encoding adapts these methods to the operations that are available when using tick-tock control.  We also apply the same SWAP patterns~\cite{Fowler2004,Szkopek2006,Stephens2009} and syndrome measurement to construct two- and three-qubit repetition codes as intermediate demonstrations towards a logical qubit.

\subsection{Linear Nearest-Neighbor Error Correction: Instruction Set and Design Rules}
Many quantum codes do not adapt well to a linear geometry; for example, topological codes cannot have a threshold in one dimension~\cite{Bravyi2009,Bravyi2013,Pastawski2015}.  Fortunately, past work has established methods for error correction in a linear or bilinear array of qubits by concatenating small codes~\cite{Gottesman2000,Fowler2004,Szkopek2006,Stephens2008,Stephens2009}, and we apply these methods to our logical qubit proposal.  Our adaptation makes some adjustments for the quantum-dot system we envision, and in the next section we introduce the tile formalism, which is a conceptual tool to aid the design and analysis of concatenated codes.  The tile formalism is a strategy for building a logical qubit using nearest-neighbor gates in a linear array of qubits, and it is based on a set of design rules that prevent some of the pathological errors that can occur in LNN circuits.

\begin{figure}
	\centering
  \includegraphics[width=8.3cm]{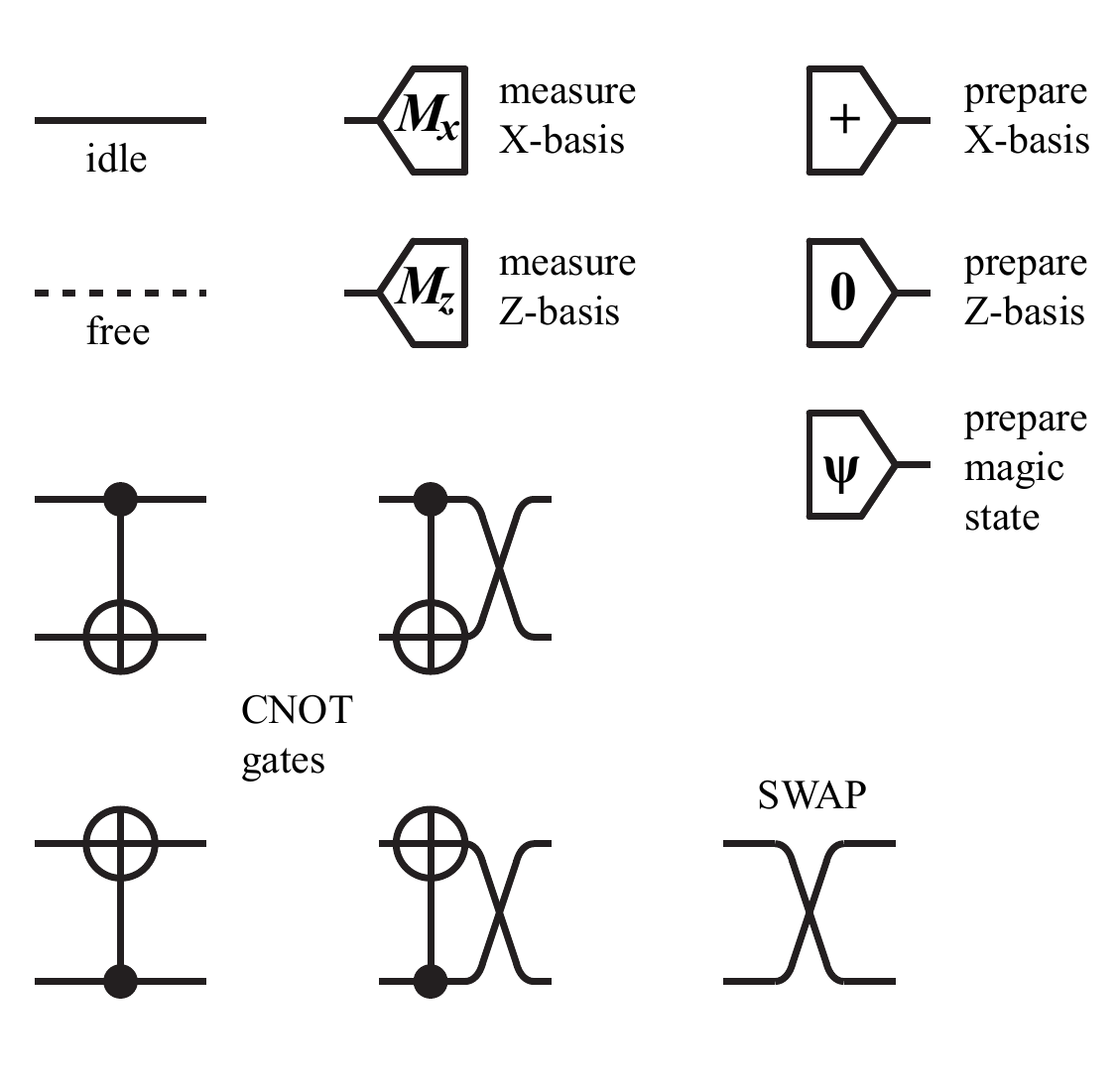}
  \caption{Standard-LNN instructions and their circuit-diagram symbols.  All of the instructions are available at the hardware level using tick-tock control, and they are native encoded operations for all CSS codes, including the codes in this proposal (note that magic state preparation is not fault tolerant and is a subroutine for distillation using other standard-LNN instructions~\cite{Bravyi2005}).  The five two-qubit gates are all combinations of CNOT gates~\cite{Nielsen2000}.  The instruction ``free'' differs from ``idle'' in that a free qubit (temporarily) has undefined state and carries no information, whereas an idle qubit does carry information.  At the hardware level, the distinction may only be a labeling, but the instructions will have different encoded representations after code concatenation.}
  \label{fig::LNN_instructions}
\end{figure}

We restrict the instructions used for error correction to a small set, which we call the standard set for an LNN architecture, or ``standard-LNN,'' shown in Fig.~\ref{fig::LNN_instructions}.  This set of instructions is closely related to CSS codes~\cite{Calderbank1996,Steane1996b}, as standard-LNN instructions are sufficient to encode, decode, and detect errors for any CSS code~\cite{Nielsen2000,Cross2009}.  Moreover, any standard-LNN encoded gate can be constructed solely from standard-LNN instructions, making this set a natural choice when using code concatenation.  The standard-LNN set consists of free, idle, preparation and measurement in both $X$ and $Z$ bases, all combinations of CNOT on two qubits (including SWAP and CNOT followed by SWAP), and magic state injection.  The instruction ``free'' is used to designate a qubit that is unused in that instruction cycle; by construction, it always follows measurement.  In contrast, idle applies to a qubit that has a defined state but is not changed in that instruction cycle.  We include the magic-state injection instruction since it is needed for universality~\cite{Bravyi2005,Knill2005}, but we consider implementations of magic-state distillation to be out of our scope.  We note that magic-state protocols based on CSS codes can be implemented by the standard-LNN set~\cite{Bravyi2005,Jones2012}.  Notably absent from this set are other Clifford gates, such as Hadamard and the $S = \sqrt{Z}$ phase gate.  However, they are not transversal in all CSS codes, nor are they needed for our encoding schemes.

The Hadamard gate is frequently included in instruction sets, so we comment briefly on how an instruction set that lacks Hadamard can still effectively implement quantum logic.  There are two common use cases for a Hadamard gate for which alternative constructions with the standard-LNN are equally efficient (or more so).  The first is to interchange $X$ and $Z$ bases after preparation or before measurement, such as in syndrome measurement circuits~\cite{Nielsen2000}.  Since preparation and measurement in both bases are in the standard-LNN set, this case is already handled.  The second common application of Hadamard is in $H$/$T$ sequences for approximating arbitrary single-qubit gates~\cite{Fowler2011,Kliuchnikov2013,Ross2014,Bocharov2015}.  The Hadamard gates can be removed by merging two consecutive Hadamard gates and the intervening $Z$-axis rotation, and then replacing the composite gate sequence with an $X$-axis rotation: $H e^{-i\theta Z} H = e^{-i\theta X}$.  The $X$-axis rotation can be generated by a magic state having a distillation protocol that is complementary to that for the $Z$-rotation magic state by converting $Z$ stabilizers to $X$ and vice versa; this is equivalent to applying transversal Hadamard to the original code, which is another CSS code that can be implemented using standard-LNN instructions.  A single remaining Hadamard gate at the beginning or end of the sequence can be implemented with magic states~\cite{Raussendorf2007}.

To make our analysis of error correction tractable, we adopt the following circuit design rules that will restrict the possible error events that can occur:
\begin{enumerate}
\item Use only standard-LNN instructions at level $L-1$ to encode all standard-LNN instructions at level $L$, beginning with tick-tock control at level 0.
\item Never perform a two-qubit gate, including SWAP, between two data qubits in the same code block.  Allowable pairs for two-qubit gates are data qubit and ancilla, two data qubits from different blocks, and in some cases two qubits in the process of encoding or decoding a block (examples discussed in next section).
\item For codes with weight-two stabilizers, use a single ancilla qubit to measure stabilizers, such that a single failure causes at most one data error.
\end{enumerate}

We have already motivated the first rule by noting that the standard-LNN set is directly related to concatenation of CSS codes.  The second rule prevents a single gate failure from introducing a weight-two error into a single code block.  The final rule similarly ensures that a single failure in the syndrome extraction circuit will introduce at most one error into a data block.  When a CSS stabilizer is measured with a single ancilla, a single failure can introduce at most a number of data errors that is half the weight of the stabilizer, rounded down~\cite{Shor1996}.  In the next section, we will describe encoding circuits for small codes using these design rules, which will simplify analysis of error propagation.

\subsection{Encoding Schemes}
We consider three closely related encoding schemes for an LNN architecture, namely the two- and three-qubit repetition codes~\cite{Shor1995,Nielsen2000} and the four-qubit subsystem code~\cite{Knill2005,Bacon2006,Stephens2009}.  These are small and simple codes that satisfy our design rules, but they can be concatenated to increase code distance.  All three codes implement the standard-LNN instruction set, making them interchangeable layers in concatenation, and they provide intermediate experiments towards a logical qubit, as described in Section~\ref{sec::experiments}.

The encoded logic gates are grouped into blocks of instructions called ``tiles,'' which provide a simple scheme for scheduling instructions to operate a logical qubit.  We can view the instructions for an LNN architecture in a two-dimensional quantum circuit diagram where the vertical dimension spans qubits and the horizontal spans time flowing to the right~\cite{Nielsen2000}.  An efficient implementation of instruction parallelism will densely fill this diagram, so we introduce interlocking tiles as a simple but effective conceptual tool for instruction scheduling.  Each tile is a sub-circuit consisting of nearest-neighbor gates on a small set of adjacent data qubits and syndrome ancillas.  We specify a tile to encode each standard-LNN instruction in each of the three codes considered here.  Note that tiles manipulate encoded states, so they only align with other tiles from the same code.

The tile formalism ensures proper logical-qubit construction, as we now explain.  The tiles naturally implement code concatenation by recursively building tiles at level $L$ from tiles at level $L-1$, where the hardware instructions are level 0. The tiles fit together perfectly in space and time, so they provide a simple method to efficiently construct concatenated LNN circuits.  Each tile satisfies the LNN design rules, ensuring that circuits composed exclusively of tiles satisfy these constraints also.  The tiles bring syndrome ancillas into contact with all data qubits for error detection.  Finally, each tile moves the ancilla qubit(s) across a code block, leaving the other side open for an interleaved two-qubit gate (described below).  Tiles provide all these features while also making instruction scheduling very simple.  Each tile has a guarantee of logical correctness, which makes it easy to verify any circuit composed of tiles.

\begin{figure}
	\centering
  \includegraphics[width=8.3cm]{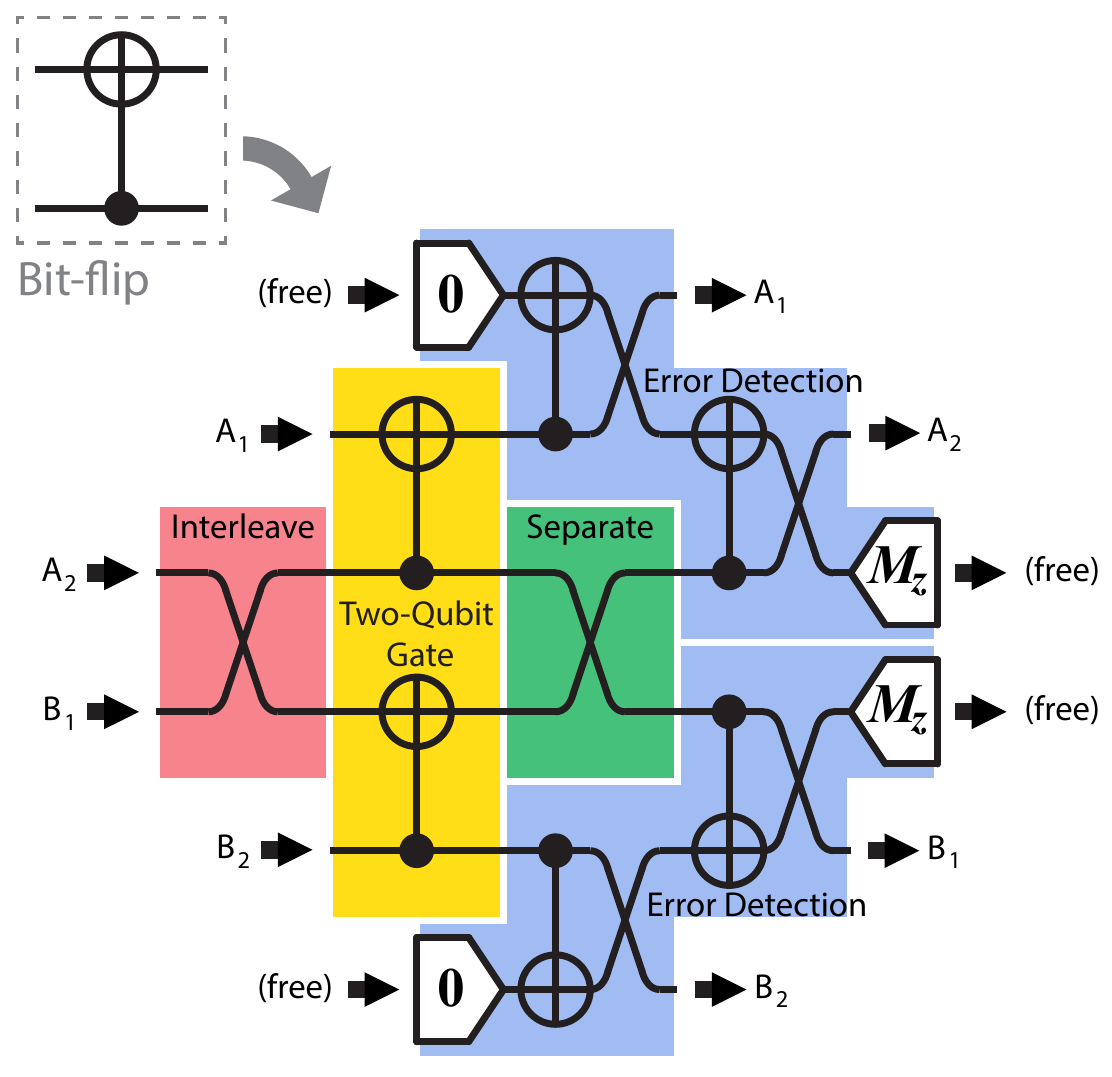}
  \caption{Tile for encoded CNOT in the two-qubit bit-flip code.  The CNOT symbol in upper left is a visual guide, and we will use ``bit-flip'' or ``phase-flip'' to clarify when necessary.  Input and output lines are labeled for convenience.  Two code blocks $A$ and $B$ have their data qubits labeled with numeric subcripts.  A core ``SWAP diamond'' of interleave, transversal CNOT, and separate operations are shown with distinct colored backgrounds.  Error detection using ancillas follows in a diagonal pass through the blocks.  These ancillas begin and end on ``free'' lines that are not encoded at the lower layer, which are at different positions before and after the tile.}
  \label{fig::CNOT_tile_rep2}
\end{figure}

The most complex circuit for an encoded standard-LNN instruction is for a two-qubit gate, so this sets the tile size for a given code.  The CNOT tile for the two-qubit, bit-flip code is shown in Fig.~\ref{fig::CNOT_tile_rep2}.  The tiles for bit-flip and phase-flip repetition codes are very similar, so we show one version of each tile and describe in words the small modification for its complement in the other code.  The tile for a two-qubit gate consists of a ``SWAP diamond'' \cite{Szkopek2006,Stephens2009} followed by error detection.  These are CSS codes, so encoded CNOT can be implemented transversally~\cite{Calderbank1996,Steane1996b,Nielsen2000}.  As shown in Fig.~\ref{fig::CNOT_tile_rep2}, data qubits from two code blocks are interleaved using SWAP gates, then a transversal CNOT is applied, then SWAP gates separate the data qubits back into their blocks.  These three steps form a diamond-shaped circuit that gives all tiles their diamond shape.  Note also that a nearly identical tile can implement any combination of encoded CNOT gates on the two code blocks (there are five such combinations), including SWAP and CNOT followed by SWAP, by modifying just the transversal operations in the middle of the SWAP diamond (shaded yellow).

Error detection is essential for a logical qubit and will be placed at the end (i.e. right side) of every tile.  Error detection is mediated by ancillas for syndrome detection~\cite{Shor1996,Steane1997,Preskill1998,Nielsen2000}, but these operations can potentially interfere with transversal two-qubit gates, which require interleaved data qubits.  The SWAP diamond primitive works best when the data blocks are adjacent, and any interspersed syndrome ancillas must be skipped over~\cite{Gottesman2000,Stephens2009}, increasing the size of the tile.  Instead, we have syndrome ancillas sweep through each data block in a diagonal, ``staircase'' circuit as in Fig.~\ref{fig::CNOT_tile_rep2}.  This sweeping action shuffles the syndrome ancilla to the other side of the block, while the data qubits move outward from the two-qubit gate just implemented.  This rearrangment is desirable because the blocks that just interacted are now positioned to interact with different neighboring blocks.  The tile in Fig.~\ref{fig::CNOT_tile_rep2} is compact, with no qubits being idle at any time.  Note also that the error detection sub-circuit in Fig.~\ref{fig::CNOT_tile_rep2} is for the bit-flip code.  The tile for CNOT in the two-qubit phase-flip code has the same interleave, transversal CNOT, and separate, but the error detection sub-circuit is different and is shown in a subsequent diagram.

\begin{figure}
	\centering
  \includegraphics[width=5cm]{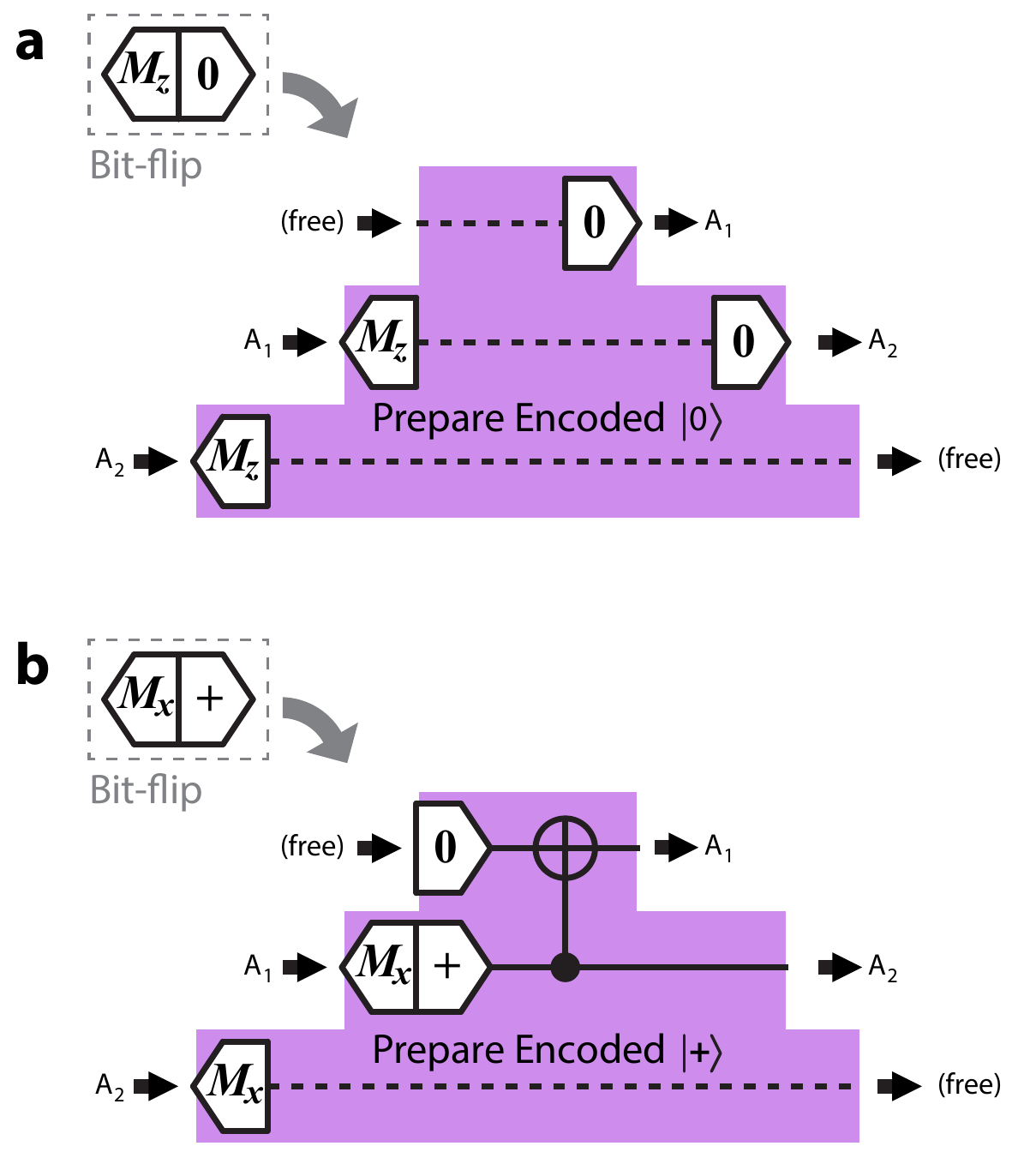}
  \caption{Half tiles for encoded measurement and preparation in the bit-flip code.   (a)~Half tile for measurement and preparation in $Z$ basis.  (b)~Half tile for measurement and preparation in $X$ basis, which requires the CNOT for encoding a $\ket{+}$ state.  In both, dashed lines denote a qubit that is free, meaning its state is unimportant (and unencoded at lower layers).  Equivalent operations in the phase-flip code are complementary, with encoded $\ket{\bar{+}}$ preparation being separable into two $\ket{+}$ preparations and $\ket{\bar{0}}$ requiring the same quarter tile as the righthand side of panel b.  Measurement operations are the same in the phase flip code (\emph{i.e.} implemented transversally).}
  \label{fig::measurement_tile_rep2}
\end{figure}

In the tile formalism, a qubit is measured and prepared in the the same time as allotted for a two-qubit gate.  Since the measure-and-prepare joint instruction acts on one encoded block, it occupies a half-tile, as shown in Fig.~\ref{fig::measurement_tile_rep2} for the two-qubit bit-flip code.  This design choice is entirely motivated by the use of code concatenation.  Although preparation and measurement may take much longer to execute than a two-qubit gate at the hardware level, an encoded two-qubit gate will be the largest tile, as it is composed of preparation, measurement, and two-qubit gates at a lower level.  After a measurement quarter tile, the constituent qubits are ``free,'' meaning they contain no quantum data and their state is temporarily unimportant (and unencoded at all lower layers).  Similarly, preparation begins with a free input line.  This can be seen in the input and output interface of the CNOT tile (Fig.~\ref{fig::CNOT_tile_rep2}); every tile for the same code implements the same interface.  In Fig.~\ref{fig::measurement_tile_rep2}a, constituent qubits are free for some number of instruction cycles since we delay preparation until required, to minimize accumulation of error.

\begin{figure}
	\centering
  \includegraphics[width=5cm]{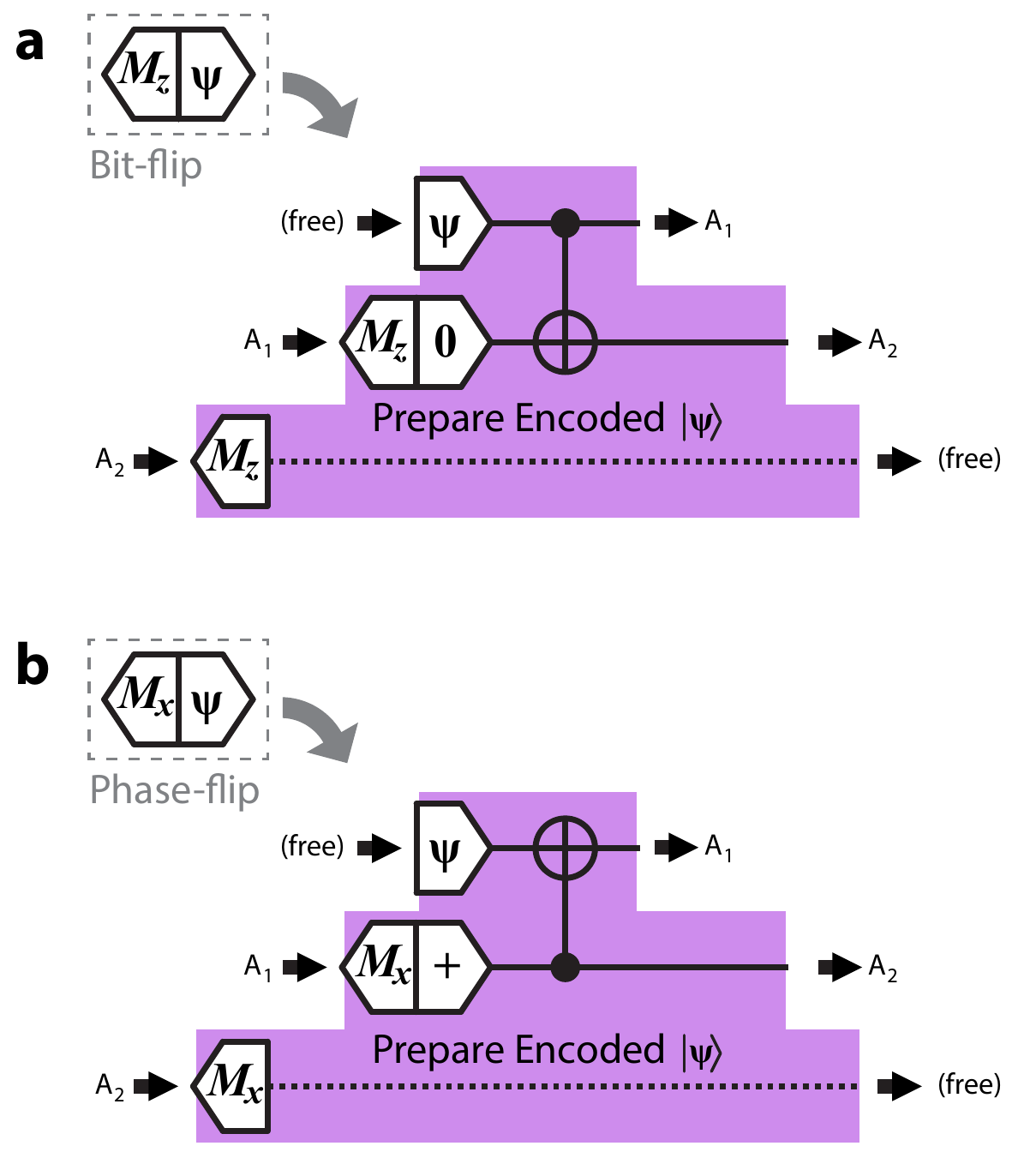}
  \caption{Half tiles for state injection in the two-qubit codes, (a) bit-flip and (b) phase-flip.  The preceding measurement is shown to complete the half tile, and the basis could be changed following Figure~\ref{fig::measurement_tile_rep2}.  Importantly, this tile alone is not fault-tolerant, because the CNOT can emit an undetected weight-two error.  This is acceptable because the injection tile is used for magic states that must be distilled.}
  \label{fig::injection_tile_rep2}
\end{figure}

\begin{figure}
	\centering
  \includegraphics[width=5cm]{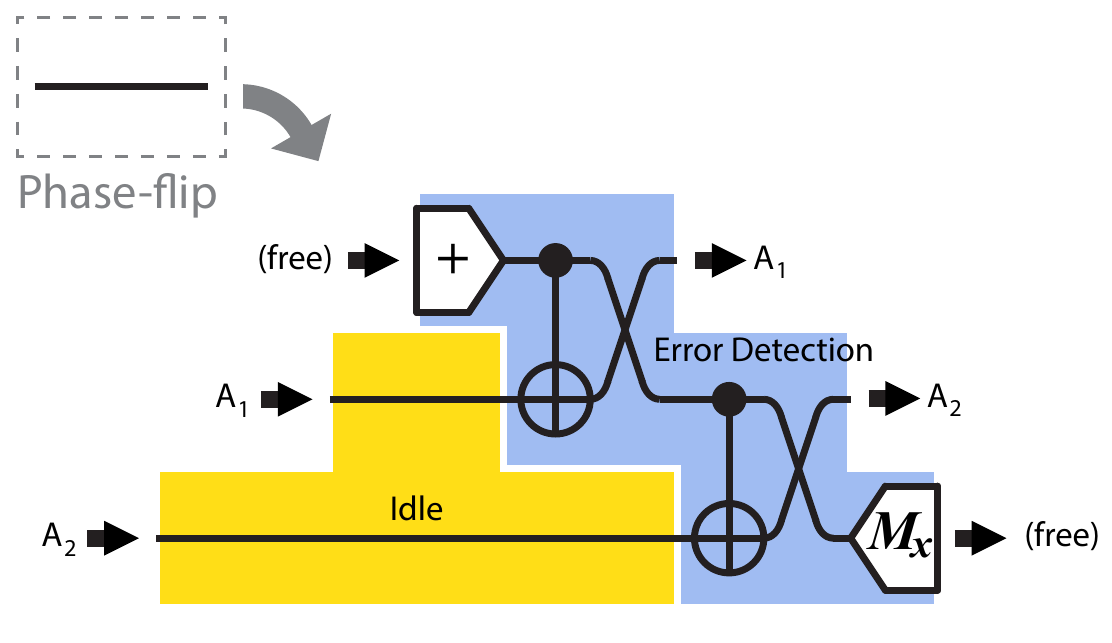}
  \caption{Half tile for idle in the phase flip code.  This shows the error-detection sub-circuit for the phase-flip code.}
  \label{fig::idle_tile_rep2}
\end{figure}

Each half-tile instruction must be matched with another half tile to form a complete diamond-shaped tile, which also determines if this mate is the code block above or below (or no block if at the edge of the linear array).  Each half tile shown in Fig.~\ref{fig::measurement_tile_rep2} is the top of a diamond, and the corresponding bottom-half tile is the mirror image about a horizontal line (not shown).  Enforcing diamond-shaped tiles enables simple scheduling without erroneous overlap of instructions.  Recall that each two-qubit gate tile has a diamond shape (Fig.~\ref{fig::CNOT_tile_rep2}; the other instruction tiles conform to this pattern.  While Fig.~\ref{fig::measurement_tile_rep2} only shows measurement and preparation in the same basis, one could also measure in $X$ basis and prepare in $Z$ basis (or vice versa) by combining the appropriate operations.  The other half tiles are state injection (Fig.~\ref{fig::injection_tile_rep2}) and idle (Fig.~\ref{fig::idle_tile_rep2}).  Figure~\ref{fig::idle_tile_rep2} also shows the error detection sub-circuit for the phase-flip code, which can be substituted into Fig.~\ref{fig::CNOT_tile_rep2} to get the CNOT tile in the phase-flip code.  Using combinations described above, this provides all of the standard-LNN encoded instructions for the two-qubit bit-flip and phase-flip codes.

\begin{figure}
	\centering
  \includegraphics[width=8.3cm]{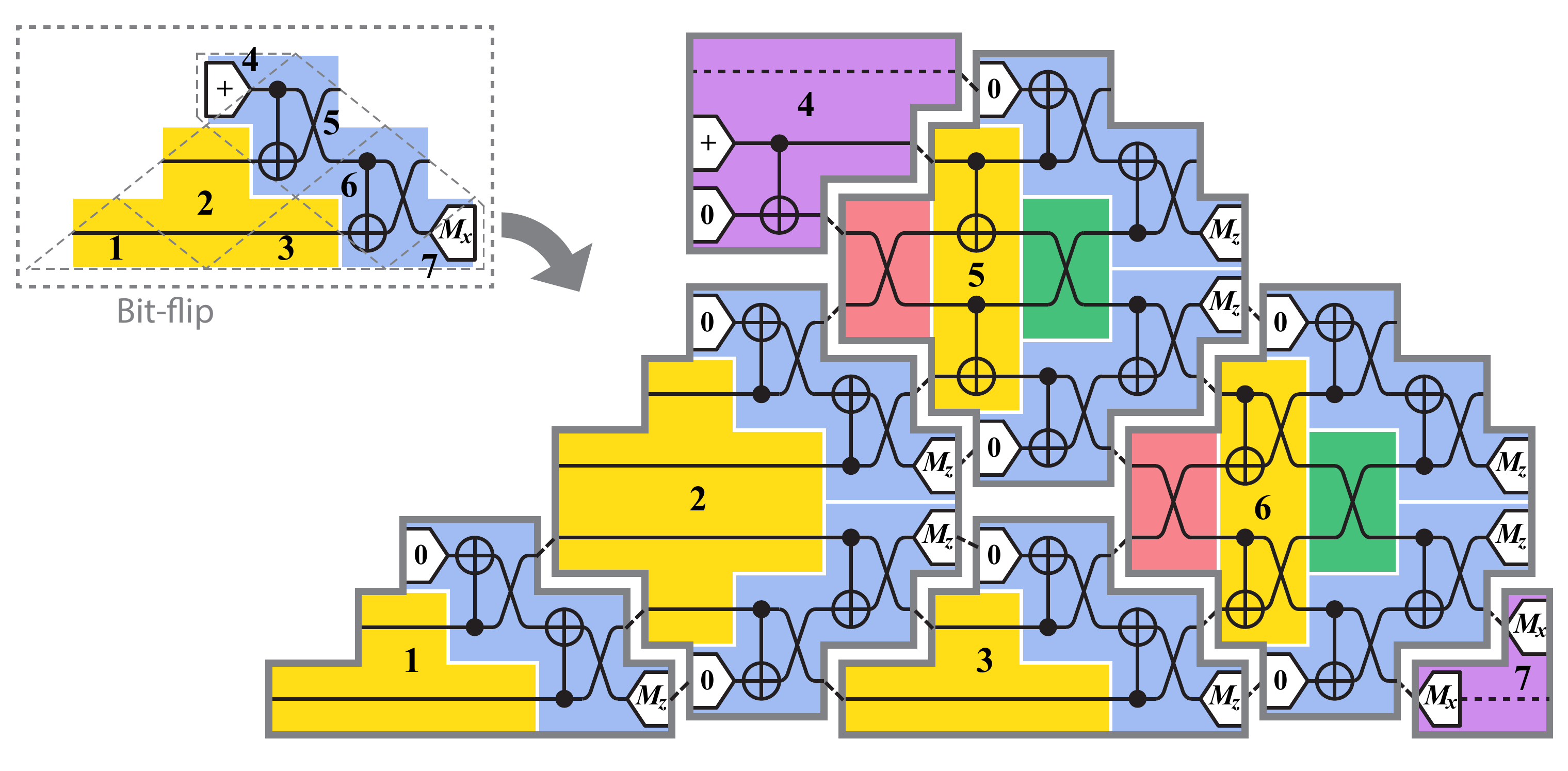}
  \caption{Concatenation of distance-two repetition codes with tiles.  A phase-flip idle tile (upper left) is encoded using the logical qubits of bit-flip codes.  The operations in the half-tile operation in upper left are divided into tile instructions with dashed lines and numbered to correspond to the bit-flip tiles at right.}
  \label{fig::concatenation_rep2}
\end{figure}

To see the advantages of using tiles for scheduling instructions, consider the circuit in Fig.~\ref{fig::concatenation_rep2} for concatenating a phase-flip code on top of a bit-flip code.  We start with a phase-flip encoded idle from Fig.~\ref{fig::idle_tile_rep2}, then replace each instruction with its appropriate tile in the bit-flip code, such as a variant of Fig.~\ref{fig::CNOT_tile_rep2} for any two-qubit gate.  In this example, we have visually separated the tiles for clarity, but they actually fit together perfectly.

Every complete tile has sufficient syndrome information to realize the full distance of the code.  For the two-qubit repetition codes and the four-qubit subsystem code (described below), this means that any single error within the tile is detected.  Correcting errors with these codes requires concatenation and syndrome processing using message passing~\cite{Knill2005,Poulin2006,Evans2008,Stephens2009,Aliferis2009}.  Appendix~\ref{app::message_passing} describes the syndrome processing that is used in Section~\ref{sec::simulation} to estimate the break-even performance of the two-qubit code.

\begin{figure}
	\centering
  \includegraphics[width=8.3cm]{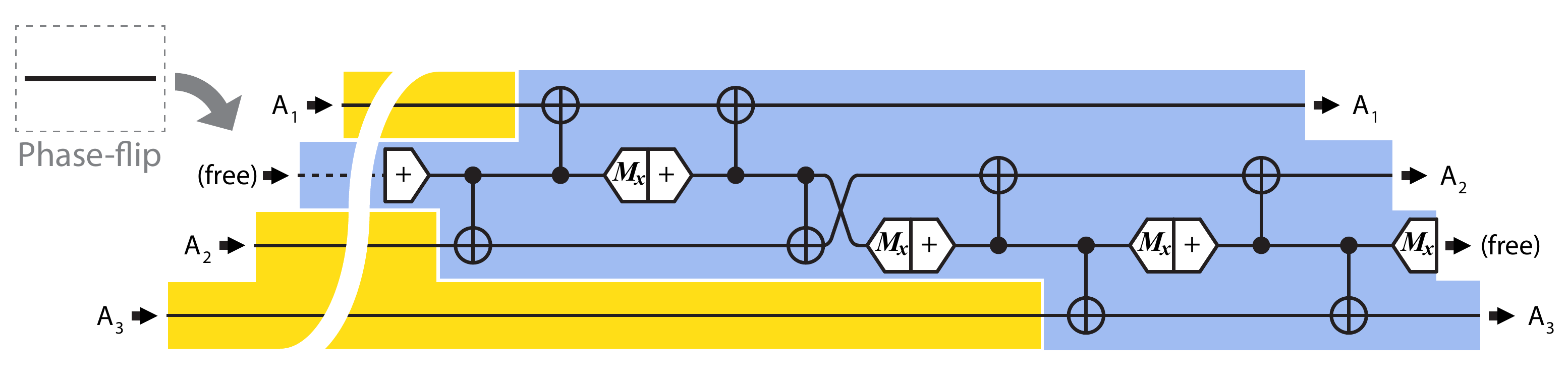}
  \caption{Idle half tile for the three-qubit repetition code with phase-flip error detection.  The figure is compressed laterally to save space, as denoted by the curved white slash.  The left side of the tile has either idle or free segments that pad width to match the CNOT tile.}
  \label{fig::idle_tile_rep3}
\end{figure}

The three-qubit repetition code is an extension of the two-qubit code, and it can detect either one $Z$ error (bit-flip encoding) or one $X$ error (phase-flip encoding).  Concatenating a bit-flip code with a phase-flip code produces Shor's nine-qubit code~\cite{Shor1995,Nielsen2000}, which can detect a single error of any type.  To identify errors with enough confidence for error correction, additional syndrome measurements are required.  This can be seen in the idle half tile in Fig.~\ref{fig::idle_tile_rep3}, where a total of four measurements are needed, two for each stabilizer generator of the code.  The redundancy in error detection circuits is needed for a tile to reliably measure the error syndrome, avoiding a scenario where a single gate failure could cause a logical error by misreading the syndrome~\cite{Shor1996,Preskill1998}.  The width of the tile in time must match the CNOT tile described below, so there are several periods of just free or idle instructions on the left side of the idle tile.  To save space in Fig.~\ref{fig::idle_tile_rep3} and subsequent figures, this waiting time is represented by a white slash across the tile.

\begin{figure*}
	\centering
  \includegraphics[width=15cm]{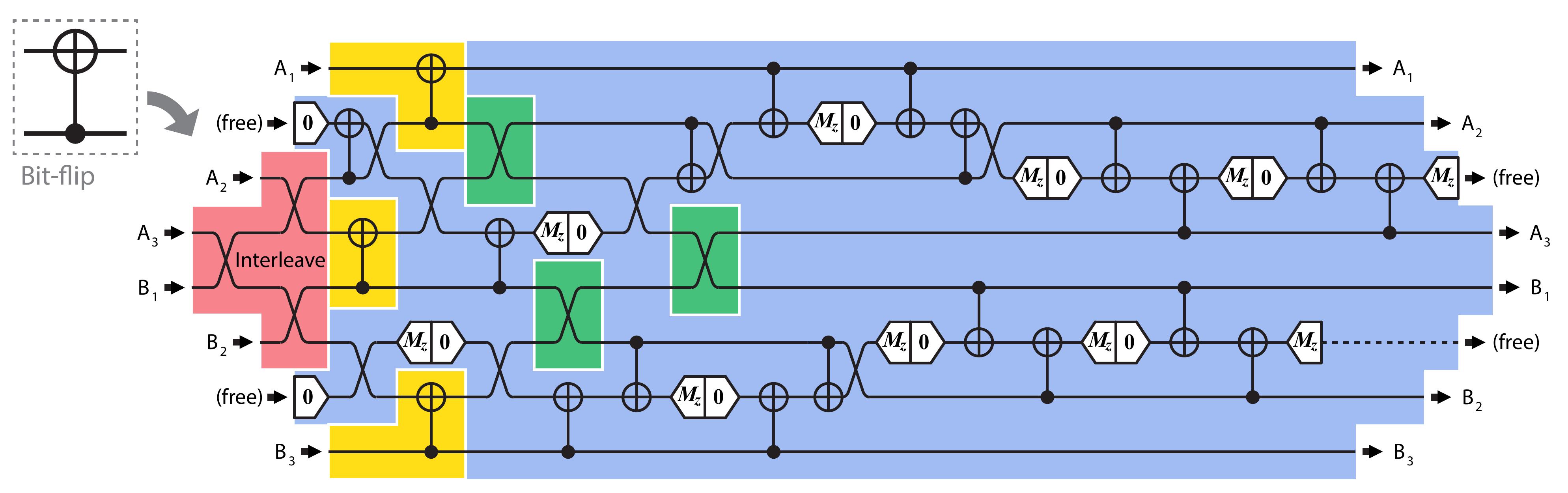}
  \caption{Encoded CNOT for the three-qubit bit-flip code.  The transversal CNOT gates (yellow background) and SWAP gates to separate the blocks (green background) are broken up by an additional error-detection circuit that is inserted after the first round of SWAP gates to catch an otherwise uncorrectable error.}
  \label{fig::CNOT_tile_rep3}
\end{figure*}

\begin{figure}
	\centering
  \includegraphics[width=6.5cm]{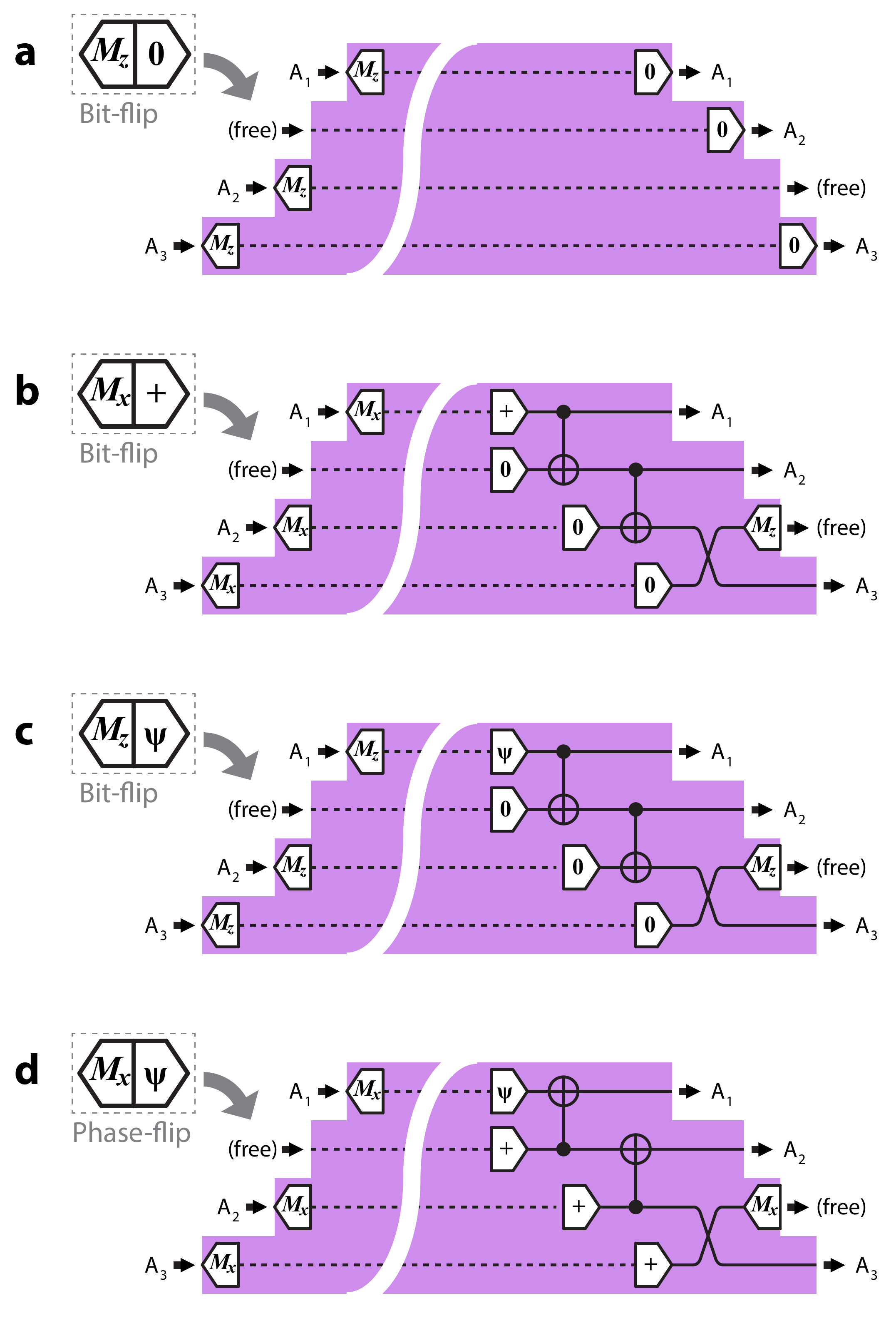}
  \caption{Half tiles for the three-qubit code, showing measurement and preparation in (a)~$Z$ basis and (b)~$X$ basis, as well as state injection for (c) bit-flip code and (d) phase-flip code.  In the phase-flip code, preparing $\ket{\bar{+}}$ is transversal, while $\ket{\bar{0}}$ uses the state injection in panel d with $\ket{\psi} = \ket{0}$.}
  \label{fig::half_tiles_rep3}
\end{figure}

The encoded two-qubit gate for the distance-three code, shown in Fig.~\ref{fig::CNOT_tile_rep3}, requires additional error detection.  Some weight-two errors produced by a single faulty SWAP gate in the interleave stage can propagate through a transversal CNOT to a weight-three error event across both blocks, which would not be correctable unless it was detected earlier.  An additional error-detection sub-circuit is inserted after interleave and before separate (shown in colored regions as before) to catch this error, though doing so displaces other operations.  This additional syndrome measurement is inserted into the block from which a logical error can propagate to the other block during the transversal operation; for the bit-flip code (shown in Fig.~\ref{fig::CNOT_tile_rep3}) this is the control block of the encoded CNOT, whereas for the phase-flip code this block is the target.  The remaining tiles for measurement, preparation, and injection are shown in Fig.~\ref{fig::half_tiles_rep3}.  As with the two-qubit code, measurement and preparation can be arranged in any combination.

The four-qubit code that we consider is effectively the same as the smallest (i.e. distance-two) Bacon-Shor code~\cite{Bacon2006,Aliferis2007,Stephens2009}, making it closely related to concatenated repetition codes.  This code has weight-four stabilizers, so syndrome measurement requires special attention to avoid introducing undetectable errors.  Before describing syndrome measurement, we will be specific about which four-qubit code we are using. It is a variant of the [[4,2,2]] code studied by Knill~\cite{Knill2005}, but only one of the logical qubits is used.  The reason for this is that designing circuits for intra-block operations between the two logical qubits and handling correlated logical errors on them are challenging problems that we leave for future work.  In keeping with the nomenclature for Bacon-Shor codes, we call the unused logical qubit a ``gauge qubit''~\cite{Bacon2006,Aliferis2007}.  The logical qubit has encoded operators $\bar{X}_L = X_1 X_2$ and $\bar{Z}_L = Z_1 Z_3$, and the gauge qubit has operators $\bar{X}_G = X_1 X_3$ and $\bar{Z}_G = Z_1 Z_2$.

\begin{figure}
	\centering
  \includegraphics[width=8.3cm]{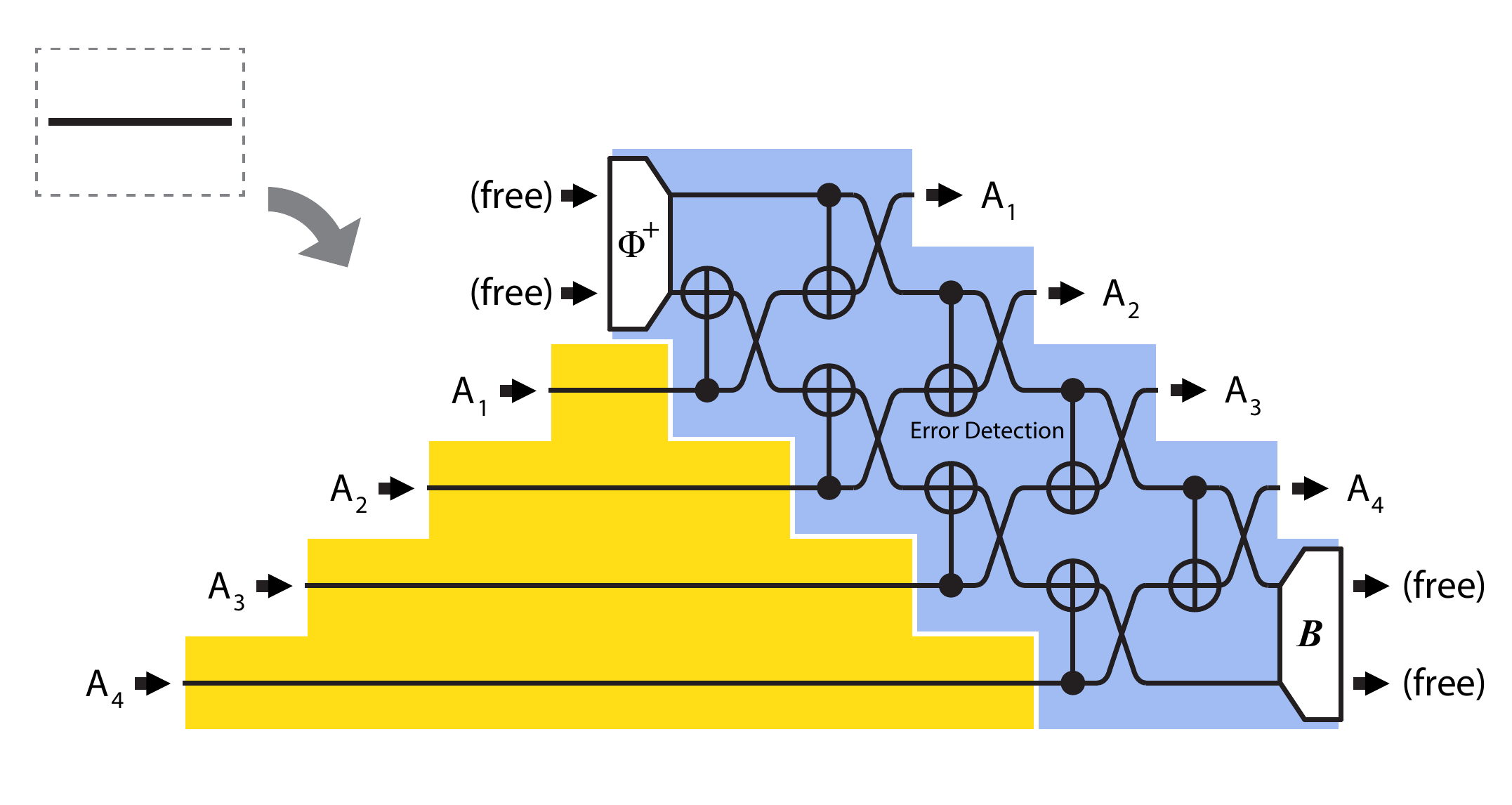}
  \caption{Idle tile for four-qubit code.  A Bell state $\ket{\Phi^+} = \left(\ket{00} + \ket{11}\right)/\sqrt{2}$ is used to measure the syndrome, because it can also detect weight-two errors introduced by the error-detection circuit.  The two-qubit ancilla is measured in the Bell basis (lower-right, denoted ``B'') to detect $X$ and $Z$ errors simultaneously.}
  \label{fig::idle_tile_four_qubit}
\end{figure}

\begin{figure}
	\centering
  \includegraphics[width=8.3cm]{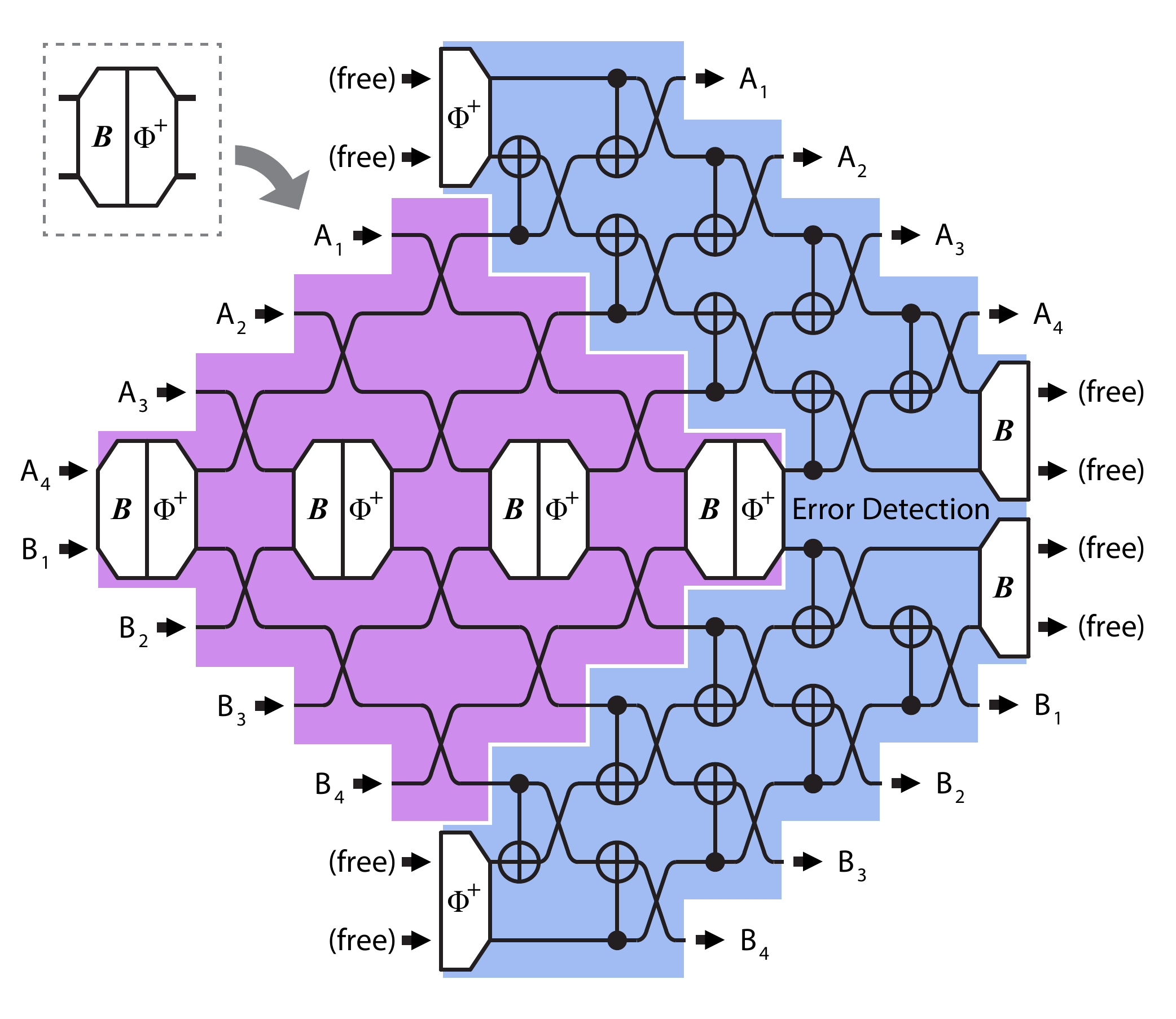}
  \caption{Tile for measurement and preparation in Bell basis.  The two operations can be implemented individually by replacing one with ``free'' instructions.  The construction produces a pair of entangled logical qubits $\ket{\overline{\Phi^+}} = \left(\ket{\bar{0}\bar{0}} + \ket{\bar{1}\bar{1}}\right)/\sqrt{2}$ and requires modified syndrome processing, described in the appendix.}
  \label{fig::Bell_basis_four_qubit}
\end{figure}

\begin{figure}
	\centering
  \includegraphics[width=8.3cm]{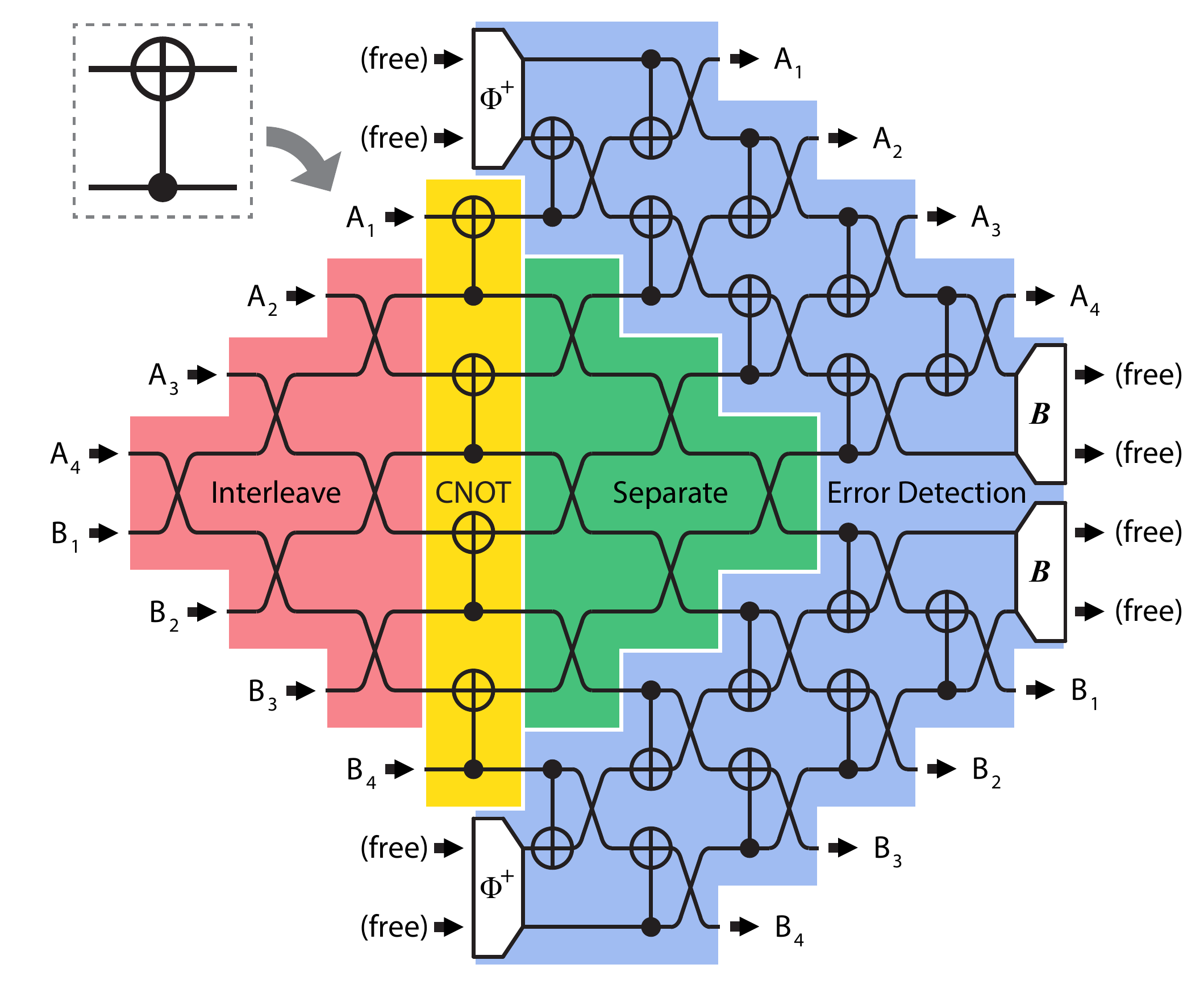}
  \caption{CNOT tile for the four-qubit subsystem code.  As with the other codes, any combination of encoded CNOTs between the two blocks can also be implemented by modifying the gate in the center of the SWAP diamond.}
  \label{fig::CNOT_tile_four_qubit}
\end{figure}

The four-qubit code has two stabilizers, $X_1 X_2 X_3 X_4$ and $Z_1 Z_2 Z_3 Z_4$.  Measuring a weight-four stabilizer has a potential problem where an error in the middle of the syndrome circuit could introduce a weight-two error that is logical and undetectable ($X_3 X_4 = \bar{X}_L$).  To solve this problem, we use a two-qubit ancilla $\ket{\Phi^+} = \left(\ket{00} + \ket{11}\right)/\sqrt{2}$ to measure both stabilizers, as shown in the idle tile in Fig.~\ref{fig::idle_tile_four_qubit}.  In addition to providing the value of both stabilizers in a Bell-state measurement, this circuit will also detect the introduction of an $\bar{X}_L$ error from a fault in the syndrome circuit.  Preparation and measurement in the Bell basis could be implemented using separate tiles for $\ket{+}$ and $\ket{0}$ preparation followed by CNOT, but a more efficient construction is shown in Fig.~\ref{fig::Bell_basis_four_qubit}; at the physical layer, Bell preparation and measurement will need to be decomposed into available hardware instructions.  The CNOT tile in Fig.~\ref{fig::CNOT_tile_four_qubit} employs the same syndrome circuit in both blocks and the SWAP-diamond shape as in previous codes.

\begin{figure}
	\centering
  \includegraphics[width=8.3cm]{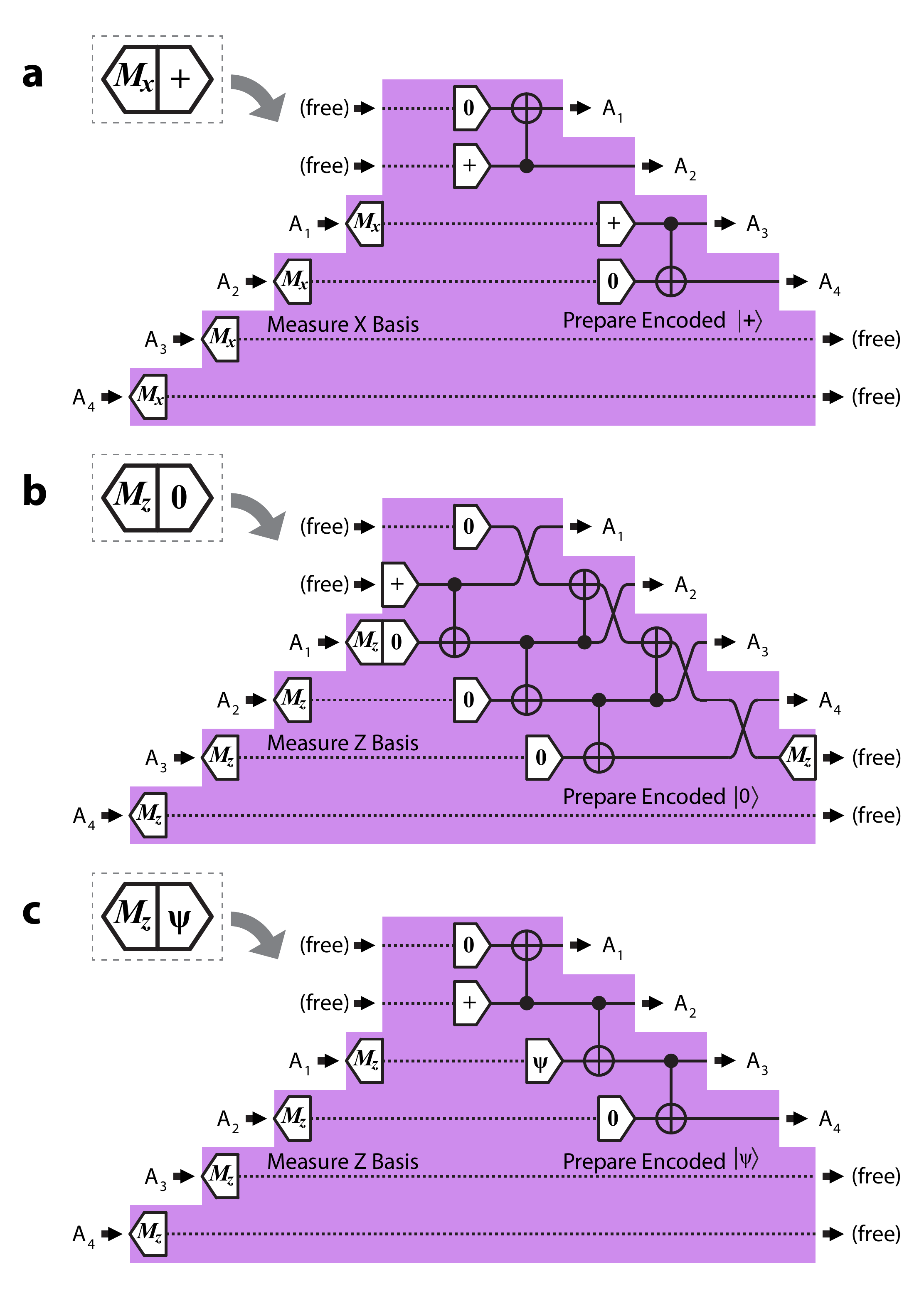}
  \caption{Half tiles for measurement and preparation in the four-qubit code, in (a) $X$ basis and (b) $Z$ basis, as well as (c) state injection.  As before, the measurement and preparation bases can be different by combining the appropriate sub-circuits.}
  \label{fig::half_tiles_four_qubit}
\end{figure}

The half tiles in Fig.~\ref{fig::half_tiles_four_qubit} are also designed to prevent an undetected logical error resulting from a single fault.  The encoded measurements are transversal and automatically fault tolerant, so we focus on the preparation circuits.  In particular, the circuits have the property that a single error from any CNOT will either be detectable or affect only the gauge qubit, just like the error-detection circuits of Figs.~\ref{fig::idle_tile_four_qubit} and~\ref{fig::CNOT_tile_four_qubit}.  Preparing $\ket{\bar{+}}$ in Fig.~\ref{fig::half_tiles_four_qubit}a is simpler than preparing $\ket{\bar{0}}$ in Fig.~\ref{fig::half_tiles_four_qubit}b, because the only weight-two errors emitted by the CNOTs in panel (a) are $\bar{X}_L$ and $\bar{Z}_G$, both of which act trivially on $\ket{\bar{+}}$.  However, any LNN CNOT between data qubits can emit an $\bar{X}_L$ or $\bar{X}_G \bar{X}_L$ error, so preparing $\ket{\bar{0}}$ requires the use of an ancilla.  The circuit in Fig.~\ref{fig::half_tiles_four_qubit}b prepares $\ket{\bar{0}}$ in a faulty way, then measures $\bar{Z}_L$ with an ancilla to detect a $\bar{X}_L$ error that could be generated by one of the CNOTs.  Consider also the injection tile in Fig.~\ref{fig::half_tiles_four_qubit}c.  The CNOTs here can emit logical errors, as described above, but this is acceptable since state injection is never fault tolerant and the magic state would need to be distilled anyway.  Finally, the idle tile in Fig.~\ref{fig::half_tiles_four_qubit}d is just idle operations on the code block followed by syndrome measurement.

We have organized our encoded instructions into tiles, because each tile is self-contained for error correction.  Each tile has sufficient syndrome information to process errors within the tile, up to the capabilities of that code.  The details of processing the syndrome are analyzed in Appendix~\ref{app::message_passing}, and the performance of the encoding schemes under standard error models is simulated in the next section.

\subsection{Simulations of Logical Qubit Performance}
\label{sec::simulation}
The performance of a logical qubit depends on the likelihood of errors and how effectively they are corrected.  In this section, we simulate some of the LNN encoding schemes of the previous section to provide performance targets for control operations.  The simulations use a simplified error model consisting of independent Pauli errors applied after every operation (including idle), following a common convention in the literature~\cite{Steane2003,Knill2005,Aliferis2006,Aliferis2007,Raussendorf2007,Cross2009,Fowler2009,Stephens2014,Stephens2014b}.  When an error occurs in a two-qubit gate, the gate is followed by one of the 15 non-identity Pauli errors with equal probability.  Although such a simplified error model cannot represent all quantum error processes, the simulations still provide guidance as to which spin-control operations require further improvement in fidelity.  As has been observed in past work, the threshold for error correction requires simulating logical error rate for several different code distances~\cite{Steane2003,Knill2005,Raussendorf2007,Aliferis2007,Evans2008,Stephens2009,Aliferis2009,Cross2009,Fowler2009,Stephens2014,Stephens2014b}.  Moreover, concatenation is necessary for distance-two codes like the two-qubit and four-qubit codes, as they can only detect errors in a single layer of encoding~\cite{Knill2005,Evans2008,Aliferis2009}.

We simulate encoding a logical CNOT ``extended rectangle''~\cite{Aliferis2006} for the two-qubit and four-qubit codes.  The three-qubit code is not shown because the matter of handling inconsistent syndrome measurements complicates both syndrome processing and how one defines an extended rectangle; to keep our scope contained, we leave detailed analysis of this code to future work.  Each simulation inserts randomly generated Pauli errors into the circuit for an encoded CNOT at one to four layers of concatenation (two-qubit code) and one to two layers (four-qubit code), which makes use of the Gottesman-Knill theorem~\cite{Nielsen2000} for efficiently simulating Clifford circuits.  The error model is depolarizing noise following every gate (or randomly negating a measurement), similar to other works in the literature~\cite{Steane2003,Knill2005,Aliferis2006,Aliferis2007,Raussendorf2007,Cross2009,Fowler2009,Stephens2014,Stephens2014b}. 

\begin{figure}
	\centering
  \includegraphics[width=\columnwidth]{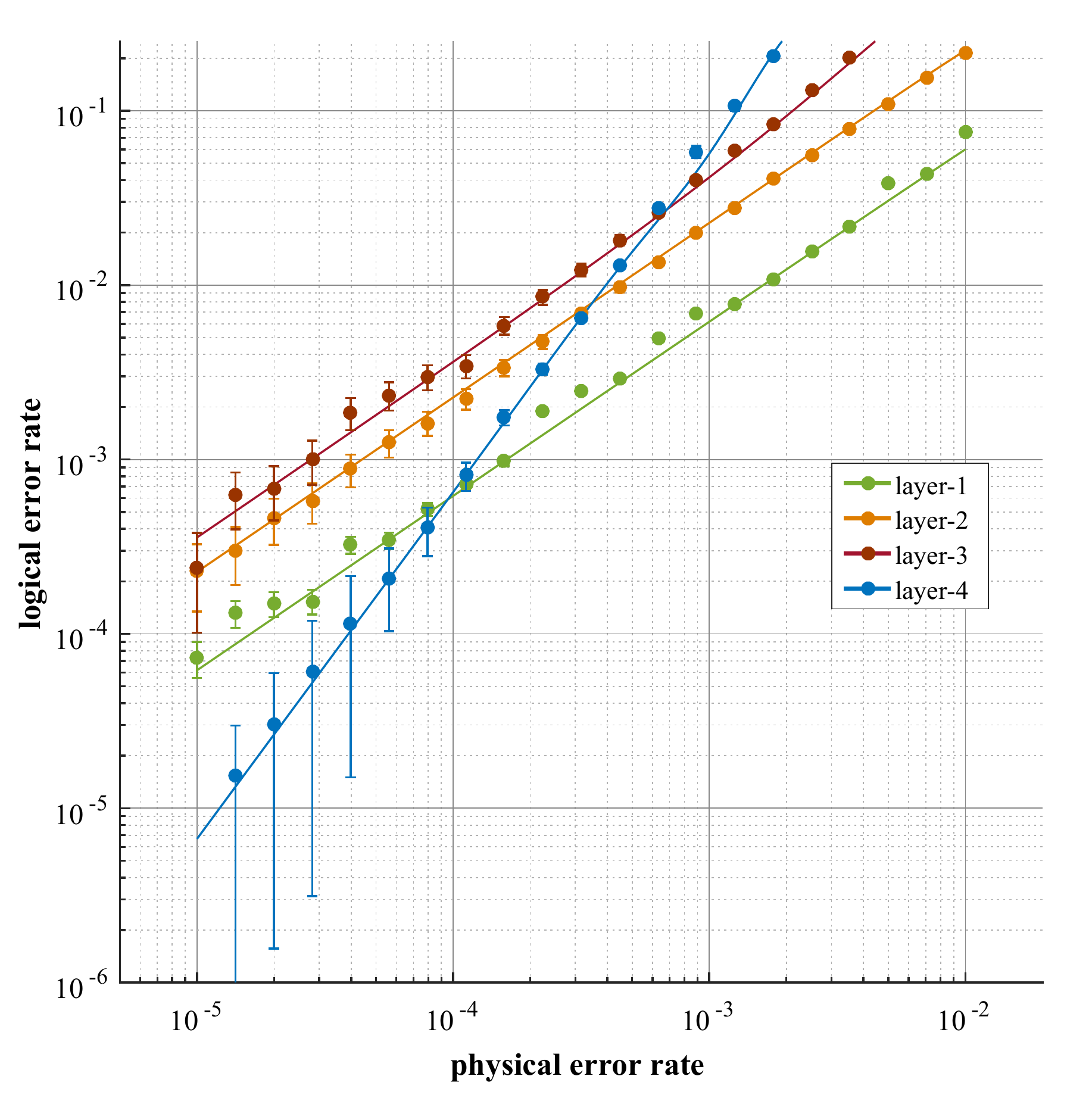}
  \caption{Simulated logical error rates for the two-qubit repetition code using concatenation.  For all traces, the sequence of concatenation alternates bit-flip, phase-flip, \emph{etc}. starting from lowest level.  Two methods of simulation are employed, Monte Carlo error generation and malignant-set counting.  The dots are Monte-Carlo generation of errors at the specified physical error rate, using the error model described in the text.  Error bars are 90\% confidence intervals estimated from the data.  The solid curves are produced by malignant-set counting or sampling, as described in the text; the logical error rate is given by Eqn.~(\ref{eqn::malignant_formula}), after the coefficients are estimated.  The correspondence between the two methods is a consistency check.  We explicitly verify that the level-four encoding corrects any single fault.}
  \label{fig::two_qubit_threshold_simulation}
\end{figure}

The results of Monte Carlo error simulations for the two-qubit code are shown in Fig.~\ref{fig::two_qubit_threshold_simulation}.  In this simulation, a logical CNOT gate is encoded in one to four layers of concatenation that alternates between bit-flip and phase-flip encoding.  Two methods of estimating logical failure rate are employed, Monte-Carlo sampling and malignant-set sampling~\cite{Cross2009,Stephens2009}.  In Monte-Carlo sampling, we generate errors independently for each gate according to a physical error parameter and count the number of logical failures.  In malignant-set sampling, we create configurations of $k$ errors and count the fraction of configurations that lead to logical failure.  The logical failure rate is then given by
\begin{equation}
\mathrm{Pr}(\mathrm{fail}) = \sum_{k = 1}^N \mathrm{Pr}(\mathrm{fail} | k \; \mathrm{errors})\mathrm{Pr}(k \; \mathrm{errors}),
\label{eqn::malignant_formula}
\end{equation}
where $N$ is the number of gates.  Since each gate has an error with the same probability $p$, the second term is simply the Bernoulli distribution, 
\begin{equation}
\mathrm{Pr}(k \; \mathrm{errors}) = \binom{N}{k} p^k (1-p)^{N-k}.
\end{equation}
The first term on the RHS of Eqn.~(\ref{eqn::malignant_formula}) is estimated by sampling from $k$-error events and determining the fraction that lead to failure, incorporating the appropriate weighting factors for the different error channels on one- and two-qubit gates, and preparation and measurement.  We explicitly verify that no single error will lead to failure in the level-four concatenated two-qubit codes.  We also truncate the sum when additional terms have no discernible affect on the plot in Fig.~\ref{fig::two_qubit_threshold_simulation}; for example, $k_{\mathrm{max}}$ is 6 for level one and 25 for level four.

These simulations suggest a threshold for the two-qubit code around $10^{-4}$.  The crossing of the level-one and level-four logical error rates occurs at $p = 9.5 \times 10^{-5}$, while the crossing of level-two and level-four curves occurs at $p = 3.1 \times 10^{-4}$.  Being more precise about the threshold would require computationally intensive simulations at higher levels of concatenation, but for now the $10^{-4}$ estimate provides a useful reference point for an LNN architecture.  

\begin{figure}
	\centering
  \includegraphics[width=\columnwidth]{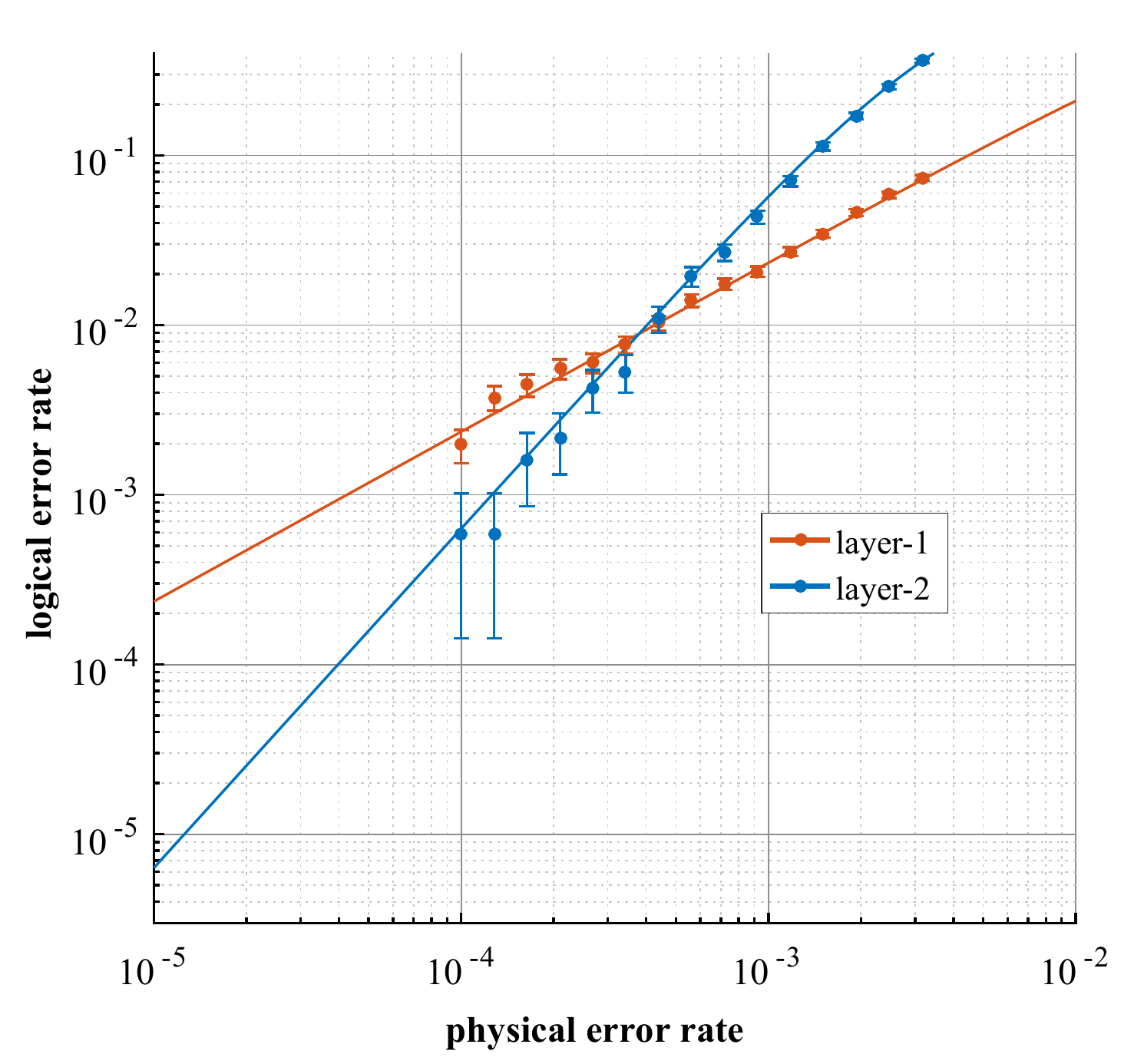}
  \caption{Simulated logical error rates for the four-qubit code using concatenation.  As in Fig.~\ref{fig::two_qubit_threshold_simulation}, both Monte Carlo error generation and malignant-set counting are employed; the dots are Monte-Carlo with error bars showing 90\% confidence intervals, and the solid curves are malignant-set sampling.  We explicitly verify that the level-two encoding corrects any single fault.}
  \label{fig::four_qubit_threshold_simulation}
\end{figure}

We also simulate the four-qubit code, as shown in Fig.~\ref{fig::four_qubit_threshold_simulation}.  The crossing of the logical-error-rate curves for layers one and two of concatenation is sometimes referred to as a ``pseudo-threshold'', which here is $3.8 \times 10^{-4}$.  It has been observed before~\cite{Stephens2009} that the threshold for further concatenation of the four-qubit code in a LNN array is slightly lower than this pseudo-threshold.  Although we leave detailed threshold simulations to future work, the results are consistent with Fig.~\ref{fig::two_qubit_threshold_simulation} in suggesting that a logical qubit in an LNN architecture would require error rates around $10^{-4}$.

We make a few comments on the resources for the encoded CNOT.  Whereas the two-qubit code requires four layers of concatenation to correct a single fault, the three- and four-qubit codes only require two layers.  Consider comparing the two- and three-qubit codes.  The tiles of the three-qubit code are larger (such as Fig.~\ref{fig::CNOT_tile_rep3}), but fewer layers of concatenation means that a distance-three encoded CNOT has about 70\% the gate count of the level-four CNOT for the two-qubit code.  With fewer error locations, we are optimistic that the three-qubit-code threshold would be similar to the two-qubit code, and perhaps the former is slightly higher.  Similarly, the CNOT in the concatenated four-qubit code requires only 20\% the number of gates as the comparable CNOT in the concatenated two-qubit code, while the pseudo-threshold in Fig.~\ref{fig::four_qubit_threshold_simulation} is very similar to that of the two-qubit code.

The experiments in Section~\ref{sec::experiments} make use of the two-qubit and four-qubit codes for intermediate demonstrations toward a logical qubit, and the simulations here provide control fidelity targets for experiments to demonstrate a ``signature'' of error correction, as explained in Section~\ref{sec::experiments}.  This signature is the characteristic quadratic dependence of logical error rate on physical rate when any single error is correctable, so failure requires two independent error events.  Figures~\ref{fig::two_qubit_threshold_simulation} and~\ref{fig::four_qubit_threshold_simulation} show that this signature can be seen even at error rates above threshold, up to $10^{-3}$ or higher, which allows an experiment to demonstrate the functionality of error correction by synthetically inserting error, even if the physical error rate is above threshold~\cite{Schindler2011,Nigg2014,Yao2012,Reed2012,Riste2015,Waldherr2014,Cramer2015}.

\section{Experimental Path to a Logical Qubit in Quantum Dots}
\label{sec::experiments}
\begin{figure*}
	\centering
  \includegraphics[width=16cm]{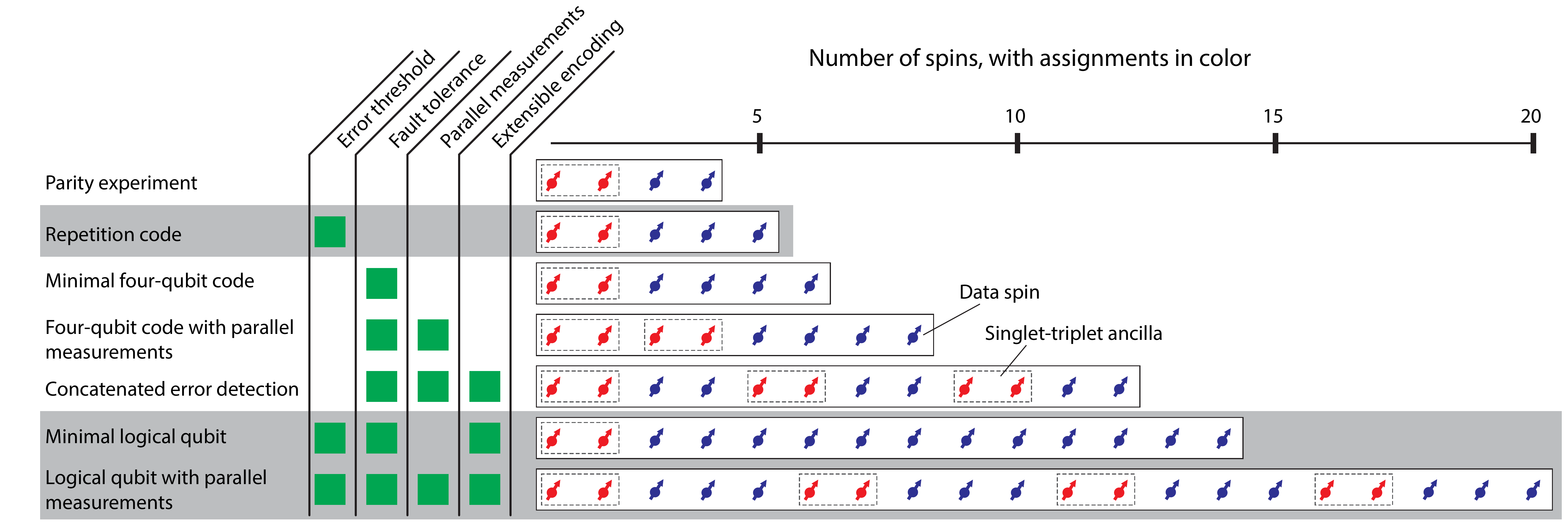}
  \caption{Summary of the experiments in this proposal.  The names of the experiments at left are grouped according to the subsections below.  The middle columns show which of the criteria for an extensible logical qubit (see Introduction) are demonstrated by the experiment, denoted with green squares.  In this context, the ``error threshold'' criterion means that one can demonstrate error correction is functioning by purposefully inserting errors, as described in the text; the ``fault tolerance'' criterion means that any single bit-flip and/or phase-flip error is detectable.  The diagrams at right are simplified device layouts showing how many coupled dots are needed for the experiment, where data spins are shown in blue and ancilla spin pairs are shown in red and grouped.}
  \label{fig::experiment_guide}
\end{figure*}

This section proposes a sequence of experiments for developing a logical qubit in quantum dots, summarized in Fig.~\ref{fig::experiment_guide}.  The experimental path demonstrates all of the requirements for an extensible logical qubit from the Introduction.  We describe the complexity of the device needed for each experiment and how the results inform the next step towards a logical qubit.  The incremental sequence of demonstrations provides numerous opportunities to improve the device design using feedback from meaningful experiments.

\subsection{Parity Measurement and Signature of Error Correction}
The parity experiment implements the two-qubit code where an ancilla detects either one bit-flip or one phase-flip error (depending on choice of encoding), such as the half tiles shown in Figs.~\ref{fig::measurement_tile_rep2} and~\ref{fig::idle_tile_rep2}.  This important experiment demonstrates the first criterion for a logical qubit, as parity measurement with an ancilla is a component of fault-tolerant error correction~\cite{Waldherr2014,Chow2014,Cramer2015,Kelly2015,Corcoles2015}.  There are two single-spin data qubits and one two-spin ancilla, requiring four dots in total.  If ancilla measurement is only available in one location in the dot array, then the ancilla will have to be swapped back to that position.  Since this device is relatively simple, it could take advantage of additional spin-control techniques that may not be extensible, such as single-spin addressed ESR and single-spin initialization~\cite{Veldhorst2014}.

The parity-measurement experiment was depicted in Fig.~\ref{fig::parity_detection}, with a prospective device layout of four dots and a charge sensor for readout.  To show that the parity-measurement process is working, one can inject errors into the qubits, as has been done in trapped ions~\cite{Schindler2011,Nigg2014}, photons~\cite{Yao2012}, superconducting qubits~\cite{Reed2012,Riste2015}, and diamond NV centers~\cite{Waldherr2014,Cramer2015}.  The first indication that parity measurement works correctly is that injected errors should predictably increase the frequency of parity-value flips.  Second, the results of measuring the individual data spins should be correlated with the parity measurements~\cite{Kelly2015}.  Finally, if the data spins are initialized as $\ket{00}$ (for bit-flip code), then the probability of observing $\ket{11}$ at the end of the experiment should be substantially suppressed when no parity flips are detected, as such an event would require two independent bit flips~\cite{Schindler2011,Reed2012,Waldherr2014,Cramer2015,Riste2015,Cramer2015,Kelly2015}.  This signature of error correction by post-selecting on not observing a parity flip can be observed in experiments that cannot demonstrate a complete logical qubit, as discussed in Section~\ref{sec::simulation}.  By initializing states that are sensitive to errors of one type (e.g. bit flip), the signature can be seen in small codes that only correct that type of error, as in several experiments below.  Similar recent theoretical work has considered experiments to show error correction is working for small surface codes with error rates near or above threshold~\cite{Tomita2014,Wootton2016}.

\subsection{Correcting One Error Type}
The parity experiment can be extended by one dot (now five dots in total) to implement the three-qubit repetition code, such as the instruction sequence in Fig.~\ref{fig::idle_tile_rep3}.  This device can both detect and correct either one bit-flip or one phase-flip error, because the two parity measurements for the three-qubit code can uniquely locate one such error.  Several recent experiments have demonstrated this encoding (or an extension of it) with an ancilla in diamond NV centers~\cite{Waldherr2014,Cramer2015} and superconducting qubits~\cite{Riste2015,Kelly2015}.

\begin{figure}
	\centering
  \includegraphics[width=8.3cm]{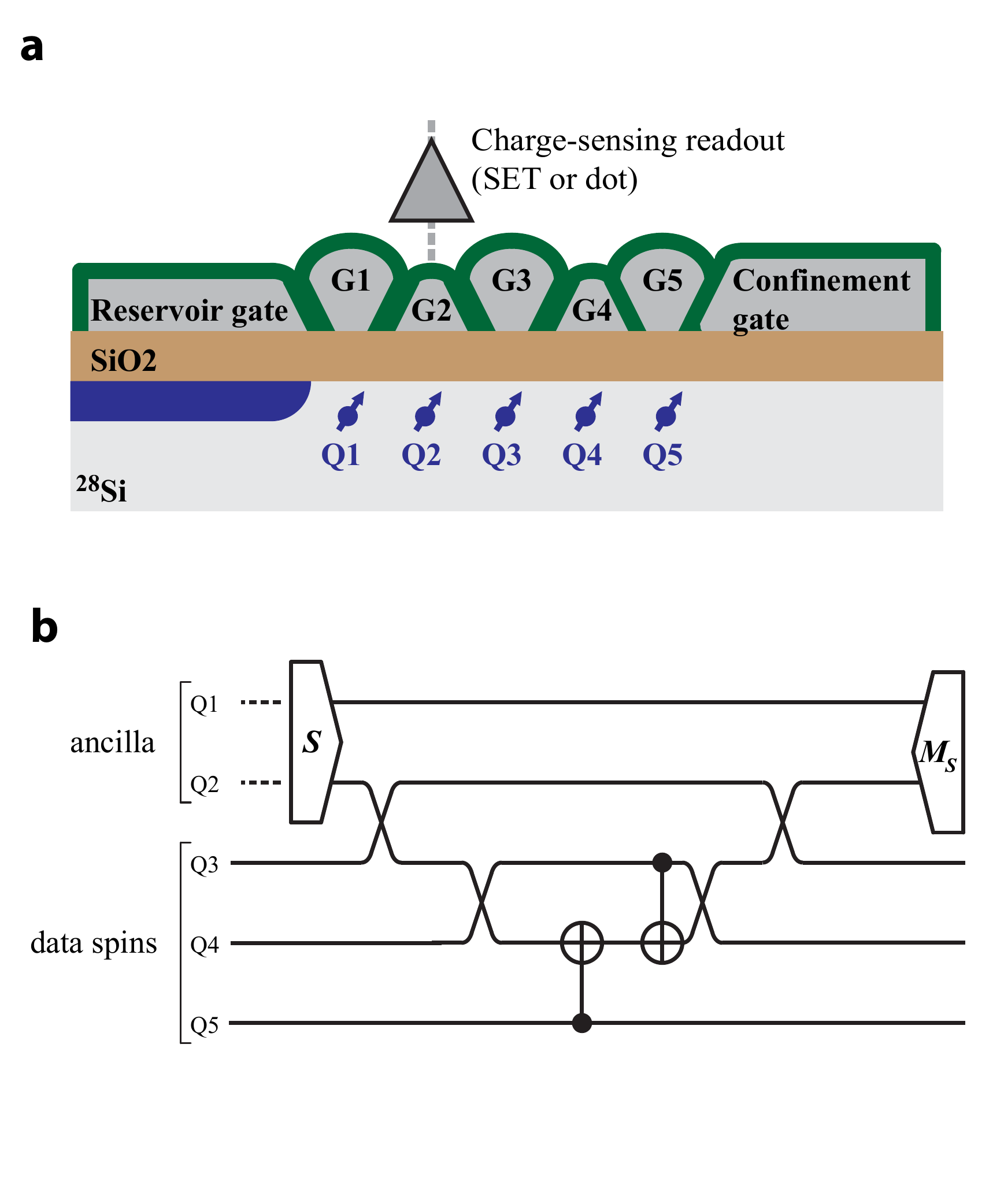}
  \caption{Device schematic and control sequence to detect one type of error, such as a bit flip.  (a)~The five-dot device has one spin-pair ancilla (Q1 and Q2) and three data spins (Q3--Q5).  (b)~Control sequence to measure the bit parity of Q4 and Q5 ($ZZ$ parity).  A complete implementation of the three-qubit code, as in Fig.~\ref{fig::idle_tile_rep3}, would require repeating this parity-detection circuit for this pair (Q4 and Q5) and the other pair of data spins (Q3 and Q4).  Each of the SWAP and CNOT operations are decomposed by tick-tock control, as shown in Fig~\ref{fig::parity_detection} and described in Section~\ref{sec::tick_tock}.  This experiment is similar to recent demonstrations in Refs.~\onlinecite{Waldherr2014,Cramer2015,Riste2015,Kelly2015}.}
  \label{fig::five_dot_experiment}
\end{figure}

The three-qubit-code experiment increases complexity by incorporating the parity measurement as a subroutine; there are now two stabilizers to measure, and each stabilizer must be measured twice.  The control sequence for the three-qubit repetition code is shown in Fig.~\ref{fig::five_dot_experiment}.  This encoding can demonstrate a logical qubit that suppresses one type of error below that of its physical qubits~\cite{Kelly2015}.  Furthermore, it is a precursor to the final logical-qubit experiment below, which concatenates the three-qubit bit-flip and phase-flip repetition codes.

\subsection{Detecting Any Single-Qubit Error}
The smallest demonstration of detecting any single-qubit data error is the four-qubit code with just one ancilla (note that the construction in Section~\ref{sec::QEC} uses two ancillas).  This implementation requires six dots, but the resulting tiles are larger to reuse the single ancilla to measure two stabilizer generators.  The new capability demonstrated by this experiment is detecting a single error of any type.  Similar recent demonstrations include stabilizing a Bell state with ancillas~\cite{Corcoles2015} and the detection of errors in an encoded state (without ancillas, however)~\cite{Zhang2012,Nigg2014}.

The next improvement to the four-qubit code is to have two measurement ancillas.  This requires eight dots, and it demonstrates both measurement parallelism and the detection of any single-qubit error.  This realizes the four-qubit code as presented in Section~\ref{sec::QEC}, such as the tiles in Fig.~\ref{fig::idle_tile_four_qubit}.

\begin{figure*}
  \centering
  \includegraphics[width=13cm]{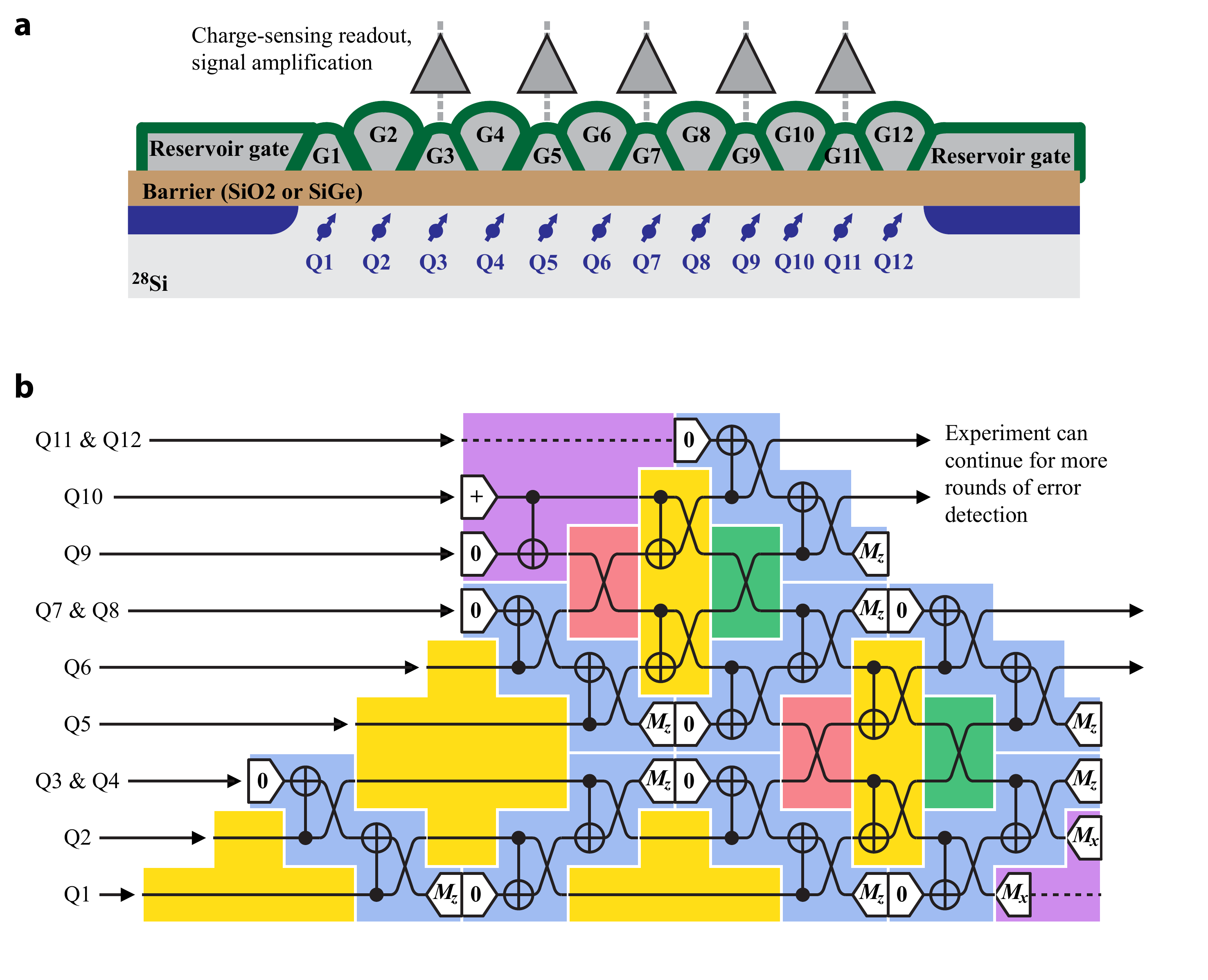}
  \caption{Experimental setup to demonstrate concatenation of bit-flip and phase-flip error-detecting codes.  (a)~Linear array of 12 quantum dots.  (b)~Diagram for code concatenation with tiles, with spins from panel a labeled on the left side.  For generality, the tiles in Section~\ref{sec::QEC} represent measurable qubits with just one line, but in tick-tock control the ancillas for measurement require spin pairs.  These spin pairs begin as (Q3,Q4), (Q7,Q8), and (Q11,Q12), and they move during the experiment following this diagram.  The spins could be initialized using techniques in Section~\ref{sec::tick_tock} to test error detection, and the experiment can be extended to more rounds by adding more tiles.}
  \label{fig::concatenation_experiment}
\end{figure*}

Using 12 dots, one can concatenate the distance-two bit-flip and phase-flip codes, as shown in Fig.~\ref{fig::concatenation_rep2}.  The control sequence for the concatenated error detection is shown in Fig.~\ref{fig::concatenation_experiment}.  Using concatenation, this experiment is essentially three copies of the parity-experiment (Fig.~\ref{fig::parity_detection}) setup integrated together.  The 12-dot experiment demonstrates two criteria for extensibility: concatenation and measurement parallelism.  The code can detect at least one error of any type, which could realize error correction if the code were concatenated again to distance four, though this experiment is outside of our scope.

These intermediate experiments on the the path to a logical qubit demonstrate several of the extensibility criteria listed in the Introduction, in increasing levels of device complexity.  The smallest four-qubit code shows the ability to detect both bit and phase errors, as well as the signature of error correction from before.  Moving to two measurement ancillas enables detection of both error types in parallel.  Finally, the concatenated two-qubit code demonstrates an extensible encoding procedure, such as encoded gates and measurements (see the tiles in Fig.~\ref{fig::concatenation_rep2}).  Collectively, the experiments to this point demonstrate all of the essential features for a logical qubit.

\subsection{Logical Qubit Demonstrations}
The logical qubit demonstrations are based on the nine-qubit code introduced by Shor~\cite{Shor1995,Nielsen2000}, which is the concatenation of the three-qubit bit-flip and phase-flip codes.  The first implementation is a minimal design that is compressed into 14 dots, comprised of nine data spins, three auxiliary data spins for an encoded block to measure the second-level syndrome, and one two-spin ancilla to measure the first-level syndrome in all four blocks.  The number of physical qubits (13) is the same as the smallest surface code~\cite{Bombin2007b,Horsman2012,Tomita2014}.  The additional idle time and SWAP operations to move this single ancilla around will penalize code performance, but the signature of error detection can be seen in the syndrome measurements even above the error-correction threshold~\cite{Schindler2011,Reed2012,Nigg2014,Waldherr2014,Riste2015,Cramer2015}.  This demonstrates many features of a logical qubit: syndrome measurement with an ancilla, code concatenation, and the ability to detect one error of any type.

The final experiment incorporates measurement parallelism, demonstrating all criteria for an extensible logical qubit.  It is another implementation of the nine-qubit Shor code using 20 dots.  Whereas the minimal logical qubit in the previous section had a single measurement ancilla shared among all data qubits in four blocks, this design is the standard implementation of code concatenation where each block has a measurement ancilla, exactly as described in the encoding tiles of Figs.~\ref{fig::idle_tile_rep3}--\ref{fig::half_tiles_rep3}.  This is four copies of the 5-dot experiment (detecting one error type) integrated together.  The distinguishing features of this experiment compared to the 14-dot implementation are that it implements the tile formalism without modification and that measurement parallelism is employed, which is crucial for extensible error correction.

\section{Discussion}
\label{sec::discussion}
We have presented a proposal that addresses all of the essential requirements for a logical qubit in silicon quantum dots.  To keep our scope contained, there are of course technology considerations that we have not analyzed in detail, which we will discuss briefly here to acknowledge their importance.  We believe that current quantum-dot technology is ready to begin developing a single logical qubit and that further improvements in materials and fabrication, such as the work cited below, will occur alongside the experimental demonstrations of Section~\ref{sec::experiments}.

The tick-tock protocol implements ESR control addressing all spins simultaneously.
The microwave power necessary to perform the global ESR control with high fidelity is dictated chiefly by the necessity to address spins which need to have different $g$-factors in order to perform fast CZ operations.  This requires the ESR pulses to be ``non-selective'' despite the significant spread in resonance frequency of the individual qubits. However, once the $g$-factor spread has been characterized and, if needed, tuned, adding more spins does not require additional power.  Heating due to ESR pulses can be mitigated by using cavities to confine the microwave modes~\cite{Abe2011,Sigillito2014,Bienfait2016}, but this presents other challenges for bringing metal electrodes to the dots.  Ongoing work in superconducting qubits for combining complex electromagnetic environments with sub-Kelvin temperatures makes us optimistic that engineering solutions are possible here as well~\cite{Dehollain2013,Gambetta2015,Brecht2016,Bejanin2016}.

The possibility of defective quantum dots is an important consideration for extensibility, so we make a few comments on this topic.  As we noted before, a single defective dot can disable a logical qubit when using LNN error correction.  For the purposes of this proposal, we believe that current technology has sufficient yield (probability of successful dot fabrication) to reach 20 coupled dots in the near term.  Schemes to handle imperfect qubit yield or qubit loss have been studied in error-correcting codes~\cite{Barrett2010,Nagayama2016}, qubit device designs~\cite{Knill2001,Ralph2005,Herrera2010,Monroe2014,Nemoto2014,Tang2016}, and quantum networks~\cite{Li2010,Munro2012,Muralidharan2014}.  Similarly, ion-trap proposals have studied how to effectively combine linear trapping regions with junctions to overcome the limitations of a strictly linear trap~\cite{Kielpinski2002,Hensinger2006,Blakestad2009,Moehring2011,Monroe2013,Sterling2014}.  In light of these methods, we expect that it is possible to arrange short linear segments of dots that meet in three- or four-way junctions, such as in Ref.~\onlinecite{Rotta2016}, enabling enough connectivity to route information around defective dots and tolerate imperfect yield.  However, developing such a scheme is outside our present scope.

We have focused our proposal on the SiMOS system, but it may certainly be adapted to other semiconductor quantum dot systems.  Confining the quantum dots in a Si/SiGe heterostructure rather than against a Si/SiO$_2$ interface may reduce the effects of disorder and charge noise, at the expense of introducing smaller valley splittings which may impair singlet initialization and measurement~\cite{Goswami2007,Kawakami2014,Eng2015,Kawakami2016}.  The controllable $g$-factor shifts in SiGe might be substantially smaller than what is observed in SiMOS dots, so the proposal might require the introduction of induced magnetic field gradients~\cite{Kawakami2014,Kawakami2016,Chesi2014,Shulman2014}.  Our scheme may also be feasible using a heterostructure based on III-V semiconductors, which have no valley degeneracy and may be engineered to have high Stark-tunable g-factor shifts~\cite{Jiang2001}.  There are inevitably large numbers of nuclear spins in III-V systems, requiring more reliance on dynamical decoupling.  Encouragingly, dynamically decoupled coherence times approaching milliseconds appear to be feasible~\cite{Malinowski2016}, although a fully hyperfine-compensating modification to our control scheme would require additional design in this case.  Finally, our scheme could be adapted to the problem of substitutional donors coupled to SiMOS-like dots or spin-shuttling channels, in which case its implementation would resemble the schemes indicated in Refs.~\onlinecite{Witzel2015,Lyon2016}.

\section{Conclusions}
We have presented a comprehensive proposal to develop a logical qubit in silicon quantum dots.  The tick-tock scheme in Section~\ref{sec::tick_tock} is an extensible way to control electron spins in quantum dots, and all of the constituent operations have recently been demonstrated with fidelity approaching the requirements of a logical qubit.  Recognizing that a linear array of exchanged-coupled dots is the most realizable device design in the near term, we have adapted simple error-correcting codes to a linear, nearest-neighbor system.  Using Monte Carlo simulations, we have estimated an error threshold of $2 \times 10^{-4}$.  Finally, we have described a sequence of experiments to demonstrate components of error correction and integrate those components into a complete logical qubit.  The final logical-qubit demonstration is a linear encoding of Shor's original quantum code, and a successful demonstration here would be a compelling argument for viability of quantum-dot logical qubits.

\section*{Acknowledgments}
We thank Menno Veldhorst, Austin Fowler, and Jason Petta for helpful insights and discussions.  M.F., A.M. and A.S.D. acknowledge support from the Australian Research Council (CE11E0001017) and the US Army Research Office (W911NF-13-1-0024).

\appendix
\section{Syndrome Processing for LNN Error Correction}
\label{app::message_passing}
This appendix describes the syndrome-processing algorithm used in the simulations of Section~\ref{sec::simulation}, where information from syndrome measurements is used to estimate the most likely configuration of errors.  The error model is a stochastic distribution of Pauli errors inserted after every control operation, idle period, preparation, and measurement~\cite{Steane2003,Knill2005}.  Syndrome processing attempts to locate and correct errors by making the most probable assignment of error conditioned on knowing the syndrome measurements.  The encoding tiles are self-contained for syndrome processing, meaning they do not share syndrome information or store it for later use.  Instead, they calculate maximum-likelihood error by searching over all error events and selecting the one with maximum probability.

\emph{Tiles are self-contained} --- To implement error correction, we first ensure that every tile can effectively detect errors up to the distance of the code~\cite{Gottesman1998,Nielsen2000}.  We say that such tiles are self-contained since they do not require syndrome information from any preceding tiles.  The two-qubit and four-qubit codes are distance-two error detecting codes, so it is only necessary that any single error is detected by the next syndrome measurement on the block.  We now list the cases to consider.  The tiles in Figs.~\ref{fig::CNOT_tile_rep2} and~\ref{fig::idle_tile_rep2} are able to detect a single bit-flip or phase-flip error event, depending on choice of encoding.  In the case of the two-qubit tile, the two blocks jointly detect errors in SWAP gates that propagate to both.  The tiles in Figs.~\ref{fig::measurement_tile_rep2} and~\ref{fig::injection_tile_rep2} have measurements that detect a single bit-flip or phase-flip error.  The circuits in Fig.~\ref{fig::measurement_tile_rep2} for preparation of $\ket{\bar{0}}$ and $\ket{\bar{+}}$ have the property that any single error is detectable by the next tile.  The state injection in Fig.~\ref{fig::injection_tile_rep2} is not fault tolerant, which is allowable since it is used to inject magic states.  The tiles in Figs.~\ref{fig::CNOT_tile_four_qubit} and~\ref{fig::half_tiles_four_qubit} have all of the same properties, with the ability to detect any single error event.

The distance-three repetition code has two additional matters to consider.  First, each stabilizer generator is measured twice, and a measurement error could cause these results to disagree.  When the stabilizer measurements are inconsistent, error correction is deferred until the next round and no corrective action is taken in that tile.  If there is an error in the data qubits, it will propagate to the next tile on that block.  Every tile can handle one incoming error on each block, so it would be the responsibility of the next tile to correct any error left uncorrected due to an ambiguous syndrome.  Second, the possibility of two errors emitted by a SWAP gate is the reason for one additional syndrome measurement \emph{before} the transversal CNOT in Fig.~\ref{fig::CNOT_tile_rep3}.  If the syndrome is consistent for both blocks and the control block for bit-flip encoding detects error (target block for phase-flip code), then there is the possibility of an undetected error in the other block.  This can occur when a SWAP gate emits two errors, one of which propagates through a transversal CNOT.  The additional syndrome measurement is needed to catch this event.

\emph{Pauli channels and message passing} --- The error model where Pauli errors occur stochastically is known as a Pauli channel~\cite{Knill2005}.  In this model, a random Pauli error can follow every standard-LNN instruction; additionally, each measurement has a probability of reporting an incorrect result.  Pauli channels are convenient because they combine to other Pauli channels, they propagate through Clifford circuits to other Pauli channels, and the conditional error channel for a stabilizer code given syndrome information is also a Pauli channel.  After correcting the syndrome, there is also a Pauli channel for the logical subspace, so there is a logical Pauli channel for every encoded standard-LNN instruction.  Hence, just as we use standard-LNN instructions to encode that same instruction set at a higher level, there is also a Pauli channel associated with that encoded instruction.  The parameters of this encoded Pauli channel are a function of the error channels for the constituent gates and the syndrome measurement outcomes.  As a result, every standard-LNN instruction has an associated Pauli channel at all levels of encoding, which is an implementation of message passing~\cite{Poulin2006}.

\begin{figure}
	\centering
  \includegraphics[width=8.3cm]{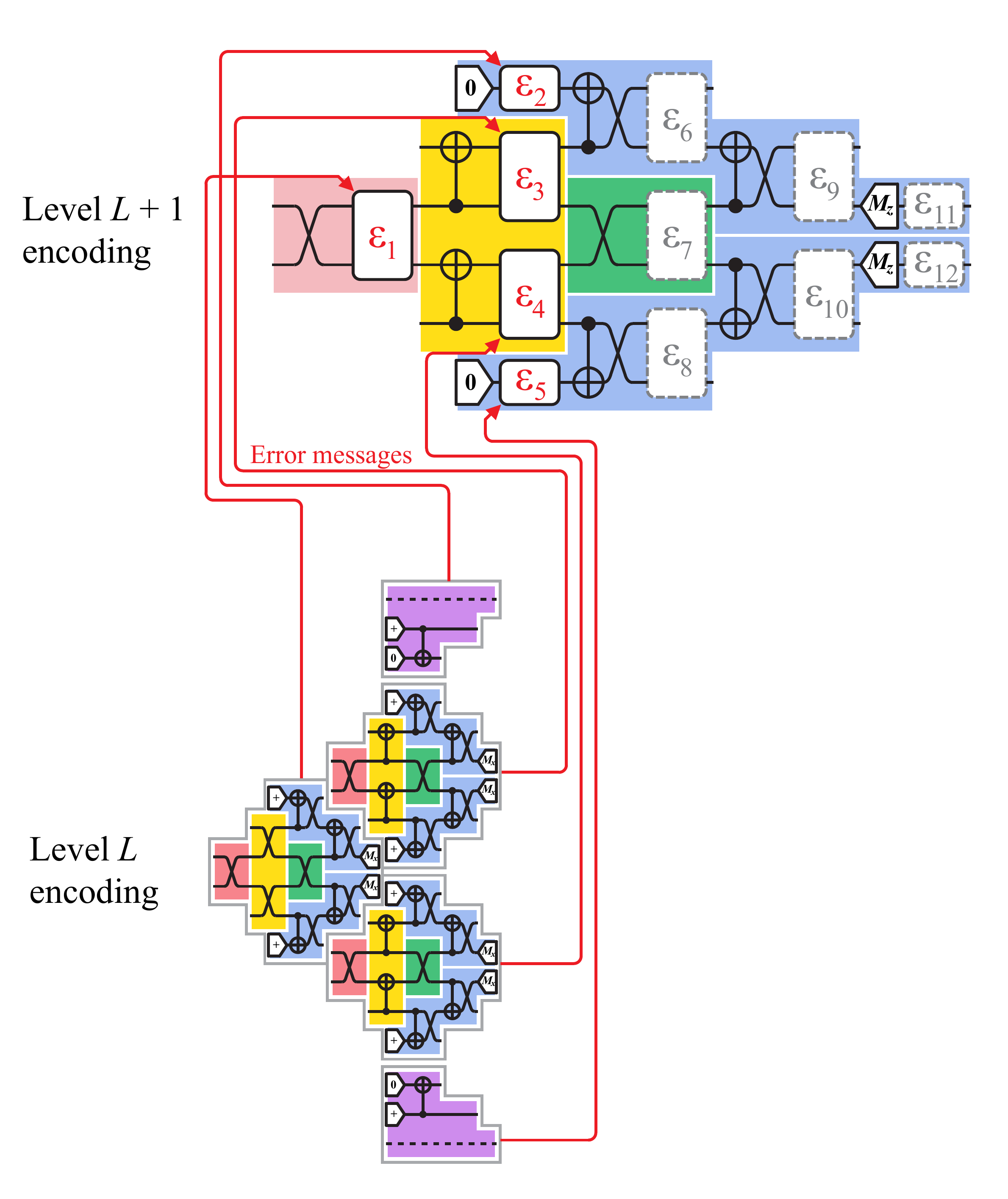}
  \caption{Message passing between layers of encoding~\cite{Knill2005,Poulin2006}, where layer $L$ is phase-flip encoded and layer $L+1$ is bit-flip encoded.  After each tile in layer $L$ processes its syndrome, it passes a message containing information on logical errors to layer $L+1$.  In this example, the first five tiles in layer $L$ have completed execution and syndrome processing; the extracted logical error channels are passed to layer $L+1$, denoted as $\varepsilon_1$ to $\varepsilon_5$.  The remaining instructions from layer $L+1$ have not executed, so the corresponding error channels are shown in grey with dashed borders.  When all 12 error channels are available, the tile at layer $L+1$ will process its syndrome and pass an error message to the next layer above.}
  \label{fig::message_passing}
\end{figure}

Message passing (MP) is a natural extension of syndrome processing in concatenated codes.  Note that in some contexts MP is known as belief propagation~\cite{Poulin2006,Leifer2008,Wang2012}, and it is a standard tool for decoding some families of classical codes~\cite{Gallager1963}.  In the original formulations of code concatenation~\cite{Preskill1998,Nielsen2000}, each layer of encoding would perform maximum-likelihood correction of errors based on syndrome measurements.  For the encoded operation, there is (at least implicitly) an error model that an encoded error occurred.  Without MP, each layer commits to a correction, which has an implicit failure probability.  With MP, each layer will still perform an assignment of error, but it also passes upwards to the next layer a measure of confidence (the message) that it made the right assignment.  This confidence measure, which could be flags for weights of errors~\cite{Knill2005,Evans2008,Stephens2009,Aliferis2009,Evans2012,Goto2013} or a Pauli channel~\cite{Poulin2006}, is based on the observed syndrome and messages from lower layers.  Consequently, the next layer above is better informed by having the message and can make a better assignment of error.  Reference~\onlinecite{Evans2012} discusses how this procedure realizes the full distance of a concatenated code.    Some form of MP is necessary for distance-two codes because they cannot assign errors in a single encoding layer~\cite{Knill2005}, and Refs.~\cite{Poulin2006,Evans2012} show that MP improves the performance of distance-three codes.  For these reasons, we employ MP in our syndrome decoding.

\begin{figure}
	\centering
  \includegraphics[width=8.3cm]{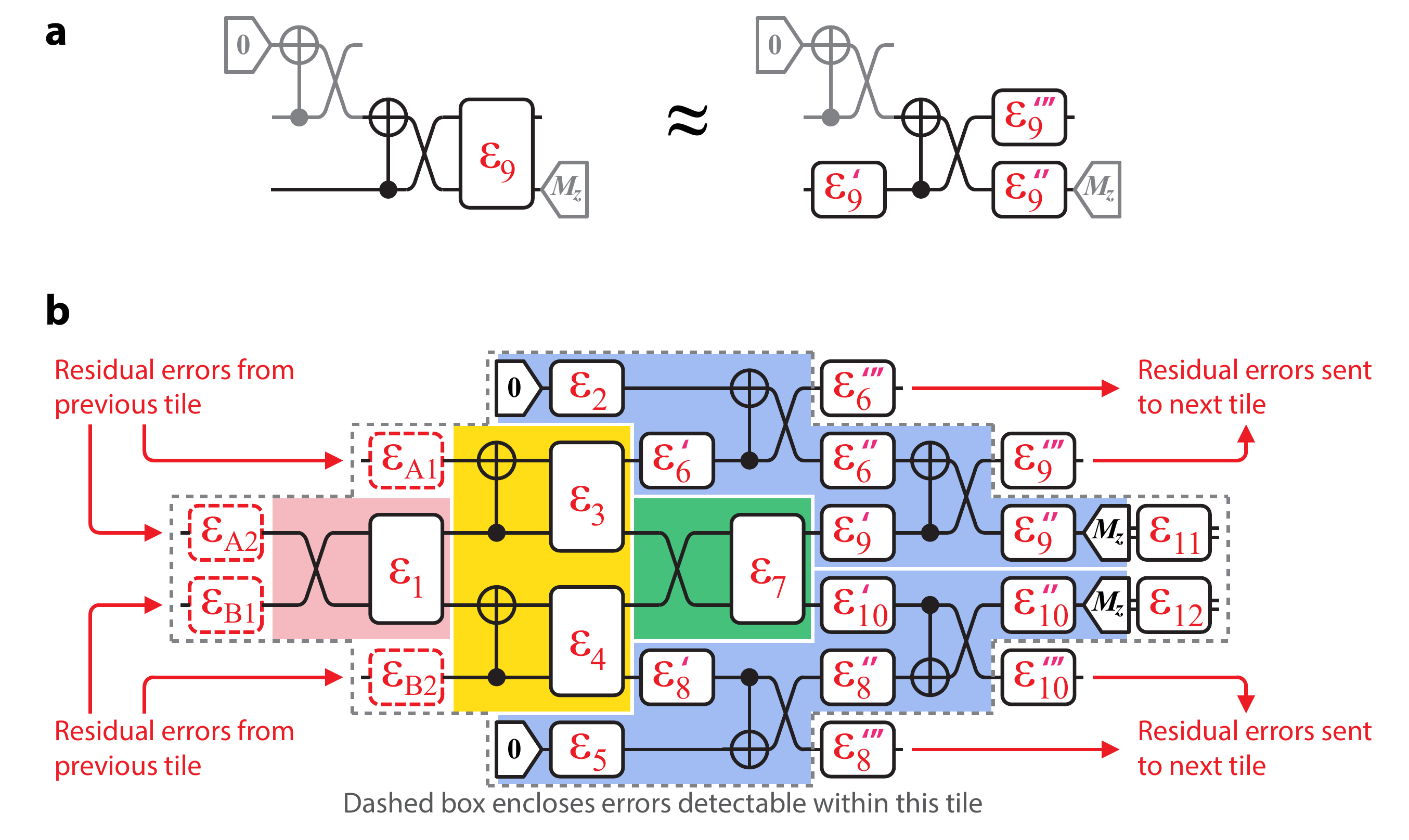}
  \caption{Error information used for syndrome processing in a tile.  (a)~Substitution to convert two-qubit error channel to single-qubit channels, separating errors that are detected by the syndrome from undetected errors.  (b)~Location of error channels after substitution in a tile that is ready for syndrome processing.  At the end (i.e. right side) of the tile, a few errors occur after syndrome processing and are not detected.  These are passed as ``residual'' errors to the next tile, just as residual errors are passed into this tile from the left.}
  \label{fig::error_splitting}
\end{figure}

\emph{Updating error likelihood using the syndrome} --- The gates within a syndrome-measurement circuit generate errors, and it is necessary to distinguish errors that propagate to measurement from those that do not.  By exploiting the stabilizer structure of the codes studied here, any two-qubit Pauli channel following a CNOT gate can be approximated by single-qubit Pauli channels before and after the gate, as shown in Fig.~\ref{fig::error_splitting}a.

To calculate error likelihood, all instructions in an encoding tile are associated with a Pauli channel.  At level~1, each Pauli channel is the assumed error model in the hardware for that instruction~\cite{Knill2005}.  At higher levels of concatenation, each Pauli channel comes as a message from syndrome processing of the tile in a lower layer.  Within a tile, syndrome processing is accomplished by searching all error events and re-weighting their probability according the observed syndrome, as follows.  Every measurement has a probability of being faulty, which is also given by the hardware error model at level~1 or a message at higher levels.  For each error event in the search, there is an anticipated syndrome result.  Depending on the observed syndrome, the probability of this error having happened is re-weighted by Bayes' theorem from elementary probability.  This probability-update procedure can be seen as a variant of the Viterbi algorithm~\cite{Forney1973}.

After re-weighting all error events, the maximum-likelihood event is selected to update the Pauli frame~\cite{Knill2005,Jones2012}.  The probabilities of other error events that would cause logical errors are combined into a logical Pauli channel that is passed as a message to the next level of encoding~\cite{Knill2005,Poulin2006,Evans2008,Stephens2009,Aliferis2009,Evans2012,Goto2013}, as depicted in Fig.~\ref{fig::message_passing}.  Searching over all error events is computationally intensive in the general case, but it is tractable for these small tiles with a finite number of error locations.  A few errors are not detectable in this tile.  As shown in Fig.~\ref{fig::error_splitting}b, these are passed as ``residual'' error channels to the next tile on the same block for processing.  Finally, after processing a two-qubit-instruction tile, the residual error channels for top and bottom blocks are correlated, and tracking such correlations would lead to an exponentially growing runaway in simulation memory; to prevent this, the distributions are ``split'' by calculating marginal distributions for the separate blocks and using this as an approximation for the closest uncorrelated joint-block distribution.

\emph{Bell-basis measurement and preparation in four-qubit code} --- The Bell-basis measurement/preparation tile in Fig.~\ref{fig::Bell_basis_four_qubit} requires some special processing of measurements.  First, note that the ``transversal'' measurement/preparation operations do not follow the familiar pattern from CNOT tiles, but rather there is a mirroring in data qubits: ($A_4$,$B_1$), ($A_3$,$B_2$), \emph{etc}.  This exploits a mirror symmetry in the code where $X_L = X_1 X_2 = X_3 X_4$ and so forth.  The reason for this change is that it prevents correlated SWAP errors from introducing a logical measurement error.  If one were to follow the typical interleave-transveral-separate pattern and replace the CNOTs in Fig.~\ref{fig::CNOT_tile_four_qubit} with Bell-state measurement/preparation instructions, the tile would be logically correct but not fault tolerant since a single SWAP failure could corrupt the encoded measurement.  The tile in Fig.~\ref{fig::Bell_basis_four_qubit} is transversal in a sense, and one can construct the logical operators $X^A_L X^B_L$ and $Z^A_L Z^B_L$ from combinations of the Bell-basis measurement on data qubits (note that superscripts ``A'' and ``B'' denote blocks).  The joint-block stabilizers $S^A_X S^B_X = \left(X^A_1 X^A_2 X^A_3 X^A_4\right) \left(X^B_1 X^B_2 X^B_3 X^B_4\right)$ and $S^A_Z S^B_Z = \left(Z^A_1 Z^A_2 Z^A_3 Z^A_4\right) \left(Z^B_1 Z^B_2 Z^B_3 Z^B_4\right)$ are also available for error detection.

The preparation circuit also merits some explanation.  The mirrored transversal preparation of $\ket{\Phi^+}$ in the data qubits creates a state that is close to the desired $\ket{\overline{\Phi^+}}$.  It is a stabilizer state with the following stabilizer generators: 
\begin{align}
& X^A_L X^B_L, \; Z^A_L Z^B_L \; & \textrm{entangled logical qubits;} \nonumber \\
& X^A_G X^B_G, \; Z^A_G Z^B_G \; & \textrm{entangled gauge qubits;} \nonumber \\
& S^A_X S^B_X, \; S^A_Z S^B_Z \; & \textrm{joint-block stabilizers.}
\label{eqn::Bell_stabilizers}
\end{align}
The missing generators for $\ket{\overline{\Phi^+}}$ are $S^A_X$ and $S^A_Z$ individually, which would also imply $S^B_X$ and $S^B_Z$ through combination with the existing joint-block stabilizers.  The syndrome-measurement circuits in Fig.~\ref{fig::Bell_basis_four_qubit} will project into a state with the same set of stabilizer operators as $\ket{\overline{\Phi^+}}$, but possibly different parity values.  The correction procedure is simple: if a stabilizer for a block is flipped, apply a corrective operation to a single data qubit, which will just be qubit 1 without loss of generality.  For example, if the measured syndrome shows $S^A_X = -\left(X^A_1 X^A_2 X^A_3 X^A_4\right)$ and $S^A_Z = \left(Z^A_1 Z^A_2 Z^A_3 Z^A_4\right)$, then we apply a corrective $Z^A_1$ (in the Pauli frame~\cite{Knill2005,Jones2012}).  Since the syndrome measurements commute with the operators in Eqn.~(\ref{eqn::Bell_stabilizers}), these operators are preserved.  In particular, the joint-block stabilizers provide error detection, so we expect $S^A_X$ and $S^B_X$ to have even parity, likewise for $Z$ stabilizers.  If odd parity is observed, then a fault in the preparation circuit has occurred, but it is detected.  Any single fault can be detected in this way, except for ``residual errors'' from the syndrome circuits, as in Fig.~\ref{fig::error_splitting}.  These events can generate at most one undetected data error per block, which can be detected by the subsequent tile.

%


\end{document}